\documentclass[preprint,floats,aps,10pt,superscriptaddress,nofootinbib,floatfix]{revtex4}
\usepackage{natbib} 
\usepackage{times} 
\usepackage{amssymb,amsmath}
\usepackage{bbold}
\usepackage{csquotes}
\usepackage{prelim2e}\usepackage[none,bottom]{draftcopy}
\draftcopyName{preprint / }{1.2} 
\ifx\pdfoutput\undefined
\usepackage[usenames,dvips]{color}
\usepackage{graphicx}     
\else
\usepackage[usenames,dvipsnames]{color}
\usepackage{epstopdf}
\usepackage[pdftex]{graphicx}
\usepackage[pdftex]{hyperref}
\hypersetup{a4paper,
    pdftitle={Manuscript on XXX} 
    pdfauthor={et al. and Uwe Thiele},  
pdfproducer={lateX},
pdfview=FitV,       
pdfstartview=FitB,
linkcolor=blue,     
citecolor=blue,     
urlcolor=red,      
breaklinks=true,    
colorlinks=true,
citebordercolor=0 0 0,  
filebordercolor=0 0 0,
linkbordercolor=0 0 0,
menubordercolor=0 0 0,
urlbordercolor=0 0 0,
pdfhighlight=/I,
pdfborder=0 0 0,   
bookmarksopen=true,
bookmarksnumbered=true
}
\fi
\ifx\pdfoutput\undefined
\DeclareGraphicsExtensions{.eps}
\else
\DeclareGraphicsExtensions{.jpg, .pdf, .tif, .png}
\fi
\graphicspath{{.},{./}} 
\usepackage[usenames,dvipsnames]{color}
\usepackage[normalem]{ulem}

\renewcommand{\vec}[1]{\mathbf{#1}}
\newcommand{\vecg}[1]{\boldsymbol{#1}}
\newcommand{\tens}[1]{\mathbf{\underline{#1}}}

\def\E{{\mathrm e}}
\def\I{{\mathrm i}}

\DeclareMathOperator{\tr}{tr}
\let\Re\relax
\DeclareMathOperator{\Re}{Re}
\let\Im\relax
\DeclareMathOperator{\Im}{Im}

\begin{document}
%
\title{Nonreciprocity induces resonances in two-field Cahn-Hilliard model}
\author{Tobias Frohoff-H\"ulsmann}
\email{t\_froh01@uni-muenster.de}
\thanks{ORCID ID: 0000-0002-5589-9397 }
\affiliation{Institut f\"ur Theoretische Physik, Westf\"alische Wilhelms-Universit\"at M\"unster, Wilhelm-Klemm-Str.\ 9, 48149 M\"unster, Germany}
\author{Uwe Thiele}
\email{u.thiele@uni-muenster.de}
\homepage{http://www.uwethiele.de}
\thanks{ORCID ID: 0000-0001-7989-9271}
\affiliation{Institut f\"ur Theoretische Physik, Westf\"alische Wilhelms-Universit\"at M\"unster, Wilhelm-Klemm-Str.\ 9, 48149 M\"unster, Germany}
\affiliation{Center for Nonlinear Science (CeNoS), Westf{\"a}lische Wilhelms-Universit\"at M\"unster, Corrensstr.\ 2, 48149 M\"unster, Germany}
%
\author{Len M. Pismen}
\affiliation{Department of Chemical Engineering, Technion - Israel Institute of Technology, Haifa 32000, Israel}
\begin{abstract} We consider a non-reciprocically coupled two-field Cahn-Hilliard system that has been shown to allow for oscillatory behaviour, a suppression of coarsening as well as the existence of localised states. Here, after introducing the model we first briefly review the linear stability of homogeneous states and show that all instability thresholds are identical to the ones for a corresponding Turing system (i.e., a two-species reaction-diffusion system). Next, we discuss possible interactions of linear modes and analyse the specific case of a ``Hopf-Turing'' resonance by discussing corresponding amplitude equations in a weakly nonlinear approach.  The thereby obtained states are finally compared with fully nonlinear simulations for a specific conserved amended FitzHugh-Nagumo system. We conclude by a discussion of the limitations of the weakly nonlinear approach. ~\\
The published version of this preprint can be found under ~\\
T.~Frohoff-Hülsmann, U.~Thiele and L.~M.~Pismen. Nonreciprocity induces resonances in two-field Cahn-Hilliard model. \textit{Phil. Trans. R. Soc. A.} 381: 20220087. 20220087. DOI: 10.1098/rsta.2022.0087
 \end{abstract}
\maketitle
%
%
%
\section{Introduction} \label{sec:intro}

Breaking Newton's third law has recently become a cherished pastime for theoretical physicists and applied mathematicians alike \cite{Pope2020l,YoBM2020pnasusa,KMSY2020pre,FHLV2021n,BFMR2022prx}. This not only formally breaks the boring symmetry in particle-particle interactions, but has dire consequences for the system's behavior: particles are not anymore attracted or repelled by their common center of mass (that remains at rest) but instead may start a chasing race as one (the ``predator'') is attracted by the other one while the latter (the ``prey'')  is repelled by the first one \cite{ChKo2014jrsi}.
In this way, oscillations and persistent motion may not only emerge for particle-based models but also characterize collective behavior as described by continuum models, e.g., nonreciprocal Cahn-Hilliard models \cite{SaAG2020prx,YoBM2020pnasusa,FrWT2021pre}, thereby providing a ``generic route to traveling states'' \cite{YoBM2020pnasusa}.\footnote{For a discussion of genericity see Ref.~\cite{FHKG2022arxiv}.}
The nonreciprocal Cahn-Hilliard model is also particularly relevant because nonreciprocal interactions keep conservation laws intact, in contrast to couplings used, e.g., in Ref.~\cite{SATB2014c}. The importance of conservation laws for pattern formation has been widely discussed: Ref.~\cite{MaCo2000n} and \cite{WiMC2005n} analyze small-scale stationary and oscillatory instabilities in the presence of a conservation law, which are relevant for pattern formation in a wide spectrum of systems, e.g., in magnetoconvection \cite{Knob2016ijam}, in crystallization of passive and active colloids \cite{TARG2013pre,OpGT2018pre}, and in the dynamics of the actin cortex of motile cells \cite{BeGY2020c,YoFB2022prl}. It has been recently shown that the standard Cahn-Hilliard model (introduced to describe phase separation in a binary mixture \cite{Cahn1965jcp,CaHi1958jcp}) also corresponds to an amplitude equation valid close to a large-scale stationary instability in systems with a single conservation law \cite{BeRZ2018pre}. This implies that reaction-diffusion systems with one conservation law, studied, e.g., in Refs.~\cite{JoBa2005prl,TeKN2005pd,IsOM2007pre,WeBK2014if,BDGY2017nc,HaFr2018np,BeGY2020c}, may be described in the vicinity of a large-scale stationary instability by a Cahn-Hilliard equation. In general, such considerations are highly relevant for the modeling of a large spectrum of biochemophysical systems ranging from proteins or the cytoplasm within biological cells to the dynamics of active colloids, microswimmers, tissues or human (or robotic) crowds \cite{KJJP2005epje,JoBa2005pb,MJRL2013rmp,RAEB2013prl,HaBF2018ptrsbs,FKLW2019rmp,SSNI2020jmb,FHLV2021n,BFMR2022prx,KrKl2022njp}. For a more extensive introduction to the nonreciprocal Cahn-Hilliard model within a wider context, see Ref.~\cite{FrWT2021pre}. Its universal importance is considered in \cite{FrTh2023preprint}.

Bifurcationally speaking, above a critical value of the nonreciprocal coupling of the two Cahn-Hilliard equations via nonequilibrium chemical potentials that keeps both conservation laws intact, oscillatory and traveling states emerge via Hopf and drift (pitchfork and transcritical) bifurcations from steady periodic or localized patterns \cite{FrWT2021pre,FrTh2021ijam}. Even though the conservation laws play an important role in the instabilities, this is similar to many other widely studied systems, e.g., reaction-diffusion models \cite{NiTU2003c,PuBA2010ap,Liehr2013,NiWa2022pd} and active phase-field-crystal models \cite{MeLo2013prl,OKGT2020c,VHKW2022msmse}. Remarkably, the nonreciprocal interactions may also result in the transformation of a stationary large-scale instability (Cahn-Hilliard instability) typical for phase separation \cite{Lang1992} into a stationary small-scale Turing-like instability with mass conservation (conserved-Turing instability). The Turing instability is well known from reaction-diffusion systems without mass conservation\cite{Turi1952ptrslsbs}. In consequence of such a transformation, in Cahn-Hilliard models nonreciprocity may cause complete suppression or arrest of coarsening \cite{FrWT2021pre} as well as the emergence of localized states \cite{FrTh2021ijam} with a slanted homoclinic snaking typical for systems with a conservation law \cite{Knob2016ijam,HAGK2021ijam}.

The original one-field passive Cahn-Hilliard model describes phase separation of binary fluid phases or isotropic solids \cite{CaHi1958jcp,Cahn1965jcp}. Eminent examples of nonvariational one-field variants include the convective Cahn-Hilliard model (broken parity symmetry) \cite{WORD2003pd,TALT2020n} and models extended by nonequilibrium chemical potentials that describe motility-induced phase separation \cite{SBML2014prl,RaBZ2019epje}. Two-field passive Cahn-Hilliard models feature a reciprocal coupling between species and are employed, e.g., to study phase separation in ternary mixtures driven by gradients of the corresponding chemical potentials \cite{Eyre1993sjam,Ma2000jpsj,NaHe2001ces,MKHK2019sm}. Nonequilibrium conditions are readily attained when two interacting number-conserving species are present. It is sufficient to make their interactions \emph{nonreciprocal} \cite{SaAG2020prx,FrWT2021pre}. Particles of one kind may be attracted by particles of another kind, while the latter may be repelled by the former. Relations of this kind naturally occur between predators and prey or between parasitic and cooperating bacteria or between catalytic particles with different phoretic response to chemical(s) produced by particles of another type. The analysis to follow reveals both parallels and differences between symmetry-breaking bifurcations in active systems with mass conservation and in reaction-diffusion systems with autocatalytic components. A combination of conserved and nonconserved species has been also considered in the framework of arrested phase separation \cite{LiCa2020jsm}. Two-field Cahn-Hilliard models with additional reaction terms in both equations (i.e., with nonvariational nonmassconserving couplings) are also widely studied, e.g.\ in \cite{OkOh2003pre,SATB2014c,ZwHJ2015pre}.

In this communication, we reconsider the nonreciprocally coupled Cahn-Hilliard model (section~\ref{sec:model}) studied in Ref.~\cite{FrWT2021pre}, and discuss it as a fully mass-conserving equivalent to the classical Turing two-species reaction-diffusion system (section~\ref{sec:linear}). Then we show that, in consequence, resonances exist between conserved-Hopf instability and conserved-Turing instability, i.e., the conserved equivalents of Hopf and Turing instability, respectively. This occurs in the vicinity of the codimension-two point where these two linear instabilities occur simultaneously (section~\ref{sec:resonance}). The modified FitzHugh-Nagumo system is considered as a  specific example in section~\ref{sec:FHN}. We close with a conclusion and an outlook in section~\ref{sec:conc}.

Mathematica notebooks that support the weakly nonlinear analysis and the codes that create all figures are published at Zenodo: http://doi.org/10.5281/zenodo.7503482

\section{Cahn-Hilliard system with nonreciprocal coupling}
 \label{sec:model}

A nonreciprocal Cahn-Hilliard (CH) model describes interactions between two species with concentrations $u(\mathbf{x},t)$ and $v(\mathbf{x},t)$ where the effective nonequilibrium chemical potential of each species depends asymmetrically on the concentration of the other species:
\begin{equation}
\mu_u = \frac{\delta{\cal F}}{\delta u}+\mu_{u}^\mathrm{nv},  \qquad  \mu_v = \frac{\delta{\cal F}}{\delta v}+\mu_{v}^\mathrm{nv},
 \label{91muas} \end{equation}
where
\begin{equation}
  {\cal F}
  = \int {\rm d}\mathbf{x}\left( \frac{\kappa_u}{2}\left|\nabla u \right|^2 + \frac{\kappa_v}{2}\left|\nabla v \right|^2 + \chi(u,v) \right)
  \label{eq:energy}
\end{equation}
is the free energy functional with the general local potential $\chi(u,v)$. The nonvariational part of the chemical potentials $\mu^\mathrm{nv}_{u}$ and $\mu^\mathrm{nv}_{v}$ (both assumed to depend on $u$ and $v$) cannot be obtained from a common functional, so that $\partial_v \mu_{u}^\mathrm{nv}\neq \partial_u \mu_{v}^\mathrm{nv}$.
Introducing the chemical potentials into the conservation laws $\partial_t u = -\nabla\cdot\vec{j}_u$ with $\vec{j}_u=-\gamma_u\nabla \mu_u$ (and similar for $v$) leads, after rescaling time and length, to the nonreciprocally coupled CH equations  
\begin{equation}
\frac{\partial}{\partial t} u= -\nabla^2 \left[\nabla^2 u + f(u,v) \right], \qquad
\frac{\partial}{\partial t} v= - \nabla^2 \left[\sigma \nabla^2 v + g(u,v)\right],
  \label{91uwas}  \end{equation}
where $\sigma = \gamma \kappa$ is the product of the ratios of mobilities $\gamma=\gamma_v/\gamma_u$ and rigidities $\kappa=\kappa_v/\kappa_u$ of the two species. Based on the functional \eqref{eq:energy}, the local terms in \eqref{91uwas} are then
  $f = - (\frac{\partial}{\partial u} \chi +\mu_{u}^\mathrm{nv})$ and $g = - \gamma (\frac{\partial}{\partial v} \chi + \mu_{v}^\mathrm{nv})$.

Note that dropping the outer Laplace operator $-\nabla^2$ from the mass-conserving system \eqref{91uwas} directly results in a typical two-species reaction-diffusion (RD) system, i.e., a system without mass conservation \cite{Pismen2006}. In the corresponding RD system, the parameter $\sigma $ represents the ratio of diffusion constants, while $f$ and $g$ represent the reaction terms. 
Below, we use this equivalence to relate the linear stability of uniform steady states of a nonreciprocal CH system directly to the linear stability of such states in RD systems. Due to mass conservation, in the CH system any homogeneous state $(u,v)=(u_s,v_s)$ automatically corresponds to a steady state. In contrast, for an RD system this requires adjustments of the constant parts of $f$ and $g$. We consider the linear stability of $(u_s,v_s)$ and show that a nonreciprocal coupling does not only allow for the classical CH instability (i.e., a large-scale stationary instability with a conservation law) but may also result in a conserved-Turing instability (i.e., a small-scale stationary instability with a conservation law) as studied in Refs.~\cite{MaCo2000n, Proc2001pla} and a conserved-Hopf instability (i.e., a large-scale oscillatory instability with a conservation law). Here, the largest available scale replaces homogeneous oscillations of a standard Hopf instability that are incompatible with conservation laws.

\section{Linear stability: Relation between conserved and nonconserved dynamics}
\label{sec:linear}

\subsection{Classification of instabilities}

\begin{table}[hbt]
\begin{tabular}{c || c | c}
& nonconserved dynamics& conserved dynamics\\
\hline
\hline
homogeneous/large-scale, stationary & Allen-Cahn (III$_\mathrm{s}$)& Cahn-Hilliard~(II$_\mathrm{s}$)\\
\hline
homogeneous/large-scale, oscillatory & Hopf~\footnotemark (III$_\mathrm{o}$)& conserved-Hopf~(II$_\mathrm{o}$)\\
\hline
small-scale, stationary & Turing (I$_\mathrm{s}$) & conserved-Turing (-)\\
\hline
small-scale, oscillatory & wave~\footnotemark  (I$_\mathrm{o}$)& conserved-wave (-)\\ 
\hline
\end{tabular}~
\caption{Naming convention of linear instabilities classified via their temporal (stationary vs.\ oscillatory) and spatial (homogeneous/large-scale vs.\ small-scale) properties for nonconserved and conserved dynamics. For reference, we give in parentheses the existing names of the instabilities in the classification by Cross and Hohenberg \cite{CrHo1993rmp}. For further explanations see main text. }
\label{tab:instab}
\end{table}
\footnotetext{Also known as ``Poincar\'e-Andronov-Hopf''.}
\footnotetext{Sometimes also called ``finite-wavelength Hopf'' or ``oscillatory Turing''.}

Before presenting the linear analysis of the model~\eqref{91uwas}, we introduce in Table~\ref{tab:instab} our classification of instabilities, for a detailed discussion see Ref.~\cite{FrTh2023preprint}.
In the literature, the Cross-Hohenberg classification~\cite{CrHo1993rmp} is often used. However, here it is not a good choice, because, in our opinion, it does not clearly distinguish between the conserved dynamics, i.e.,~the model~\eqref{91uwas}, and the nonconserved dynamics, i.e.,~the corresponding RD system (Eq.~\eqref{91uwas} without the leading $-\nabla^2$). In Table~\ref{tab:instab} we distinguish two main classes - nonconserved and conserved dynamics, each divided into four subclasses depending on the spatial and temporal character of the growing modes near the onset.

First, if the imaginary part of the temporal eigenvalue is zero, the unstable mode grows monotonically - the instability is stationary, if not, it defines the temporal frequency of the oscillation - the instability is oscillatory. Second, the wavenumber encoding the spatial structure of the unstable mode at onset is either zero or finite. In the former case one has an homogeneous or large-scale instability, in the latter the wavenumber defines the characteristic length scale of a small-scale instability.\footnote{Small-scale and large-scale instability are also referred to as ``short-wave'' and ``long-wave'' instability, respectively. Alternatively, but much less frequently, also ``short-scale'' and ``long-scale'' instability is used \cite{GoNP1994pf}.}
In the nonconserved case, the instability at $k=0$ always corresponds to a homogeneous (or global) mode, as each point of a finite or infinite domain grows monotonically or oscillatory without any spatial modulation. Thus, we refer to it as a ``homogeneous instability''. Such an homogeneous behavior is, however, incompatible with a fully conserved dynamics.\footnote{This is the case for model~\eqref{91uwas} where all components are conserved. For systems with fewer conserved quantities than dynamically evolving fields, homogeneous oscillations compatible with the conservation law are possible. In such systems one may also encounter Turing instabilities of nonconserved modes that have then to be considered in conjunction with also existing neutral modes due to the conserved quantities.} Instead, the stationary or oscillatory mode with the smallest wavenumber compatible with the boundary conditions is excited, e.g.~for periodic boundary conditions its wavelength equals the domain size. This is called a conserved-Hopf (or oscillatory long-wave) instability. In the following, we address all linear instabilities by the names given in Table~\ref{tab:instab}.
\subsection{Instability thresholds}
\label{sec:model2}
Using the ansatz $(u,v)=(u_s,v_s) + \varepsilon (u_1,v_1)\,\exp(\lambda t+ \text{i} \vec{k}\cdot\vec{x})$ with $\varepsilon\ll1$, and abbreviating partial derivatives with respect to $u$ and $v$ as subscripts, e.g.~$\frac{\partial}{\partial u} f = f_u$, the linearized equations~\eqref{91uwas} are
 \begin{equation} 
   \left(\tens{L}(k^2)-\lambda \tens{\mathbf{1}}\right)\,\left(\begin{array}{c}u_1\\v_1 \end{array} \right)=\vec{0}
  \quad\mathrm{with}\quad  
   \tens{L}(k^2)= k^2\left( \begin{array}{cc} f_u- k^2 & f_v \\  g_u & g_v-\sigma k^2
  \end{array} \right) = k^2 \,\widetilde{\tens{L}}(k^2)\,.
\label{91jac}
\end{equation} 
The derivatives in the Jacobian matrix $\tens{L}$ are computed at the homogeneous state $(u_s,v_s)$ which we do not need to specify. The eigenvalues are given by
\begin{equation}
 \lambda= k^2 \tilde\lambda \qquad\mathrm{with}\qquad  \tilde\lambda=\frac{\tr \widetilde{\tens{L}}}{2} \pm \sqrt{\frac{(\tr \widetilde{\tens{L}})^2}{4} - \det \widetilde{\tens{L}}}\,.
 \label{eq:lambda}  \end{equation}
It is important to note that the expression \eqref{91jac} differs from the classical Turing problem of the linear stability of a two-component RD system \cite{Turi1952ptrslsbs,Pismen2006} solely by the factor $k^2$. This implies that $\det\tens{L}(k^2)$ has zeros wherever $\det\widetilde{\tens{L}}(k^2)$ does.  Moreover, since thresholds of symmetry-breaking instabilities correspond to zero crossings of maxima of $\mathrm{Re}\lambda(k^2)$ (where its derivative with respect to $k^2$ vanishes),  whenever both $\mathrm{Re}\tilde\lambda(k^2)=0$ and $\partial_{k^2}[\mathrm{Re}\tilde\lambda(k^2)]=0$,
also $\mathrm{Re}\lambda(k^2)=0$ and $\partial_{k^2}[\mathrm{Re}\lambda(k^2)]=\partial_{k^2}[k^2\mathrm{Re}\tilde\lambda(k^2)]=0$, implying identical instability thresholds in the conserved and nonconserved case.\footnote{The determinant $\det\tens{L}(k^2)$ has then, in addition to the zeroes of  $\det\widetilde{\tens{L}}(k^2)$, a persistent zero at $k=0$. This may be irrelevant for linear stability but is important for weakly nonlinear analysis.} Therefore, the stability diagrams for homogeneous states of model~\eqref{91uwas} and of the corresponding RD model are identical. The discussed equivalence directly implies that the product of the ratios of mobilities and rigidities in the nonreciprocal Cahn-Hilliard system (conserved case) $\sigma$ takes the role of the ratio of diffusion constants in the corresponding RD system (nonconserved case). However, besides their zero crossings, the dispersion relations in a conserved ($\lambda(k)$) and nonconserved ($\tilde\lambda(k)$) case are different, and distinctions between the two cases are important for nonlinear analysis and detection of secondary instabilities. 

The instability thresholds for a conserved  system can therefore be established by analyzing the eigenvalues $\tilde\lambda$ for the nonconserved case. The onset of all stationary  instabilities, i.e., with $\Re \tilde\lambda = \Im \tilde\lambda = 0$, is determined by $\det \widetilde{\tens{L}}=0$, i.e.,
\begin{align}
 0 =f_u g_v - f_v g_u - k^2(\sigma f_u +g_v)+\sigma k^4 \equiv A-k^2B+\sigma k^4,
 \label{91Jas}  \end{align}
which gives the following wavenumbers of marginally stable modes 
\begin{equation}\label{eq:kpm}
k_\pm^2 = \frac{B}{2\sigma}\left[ 1 \pm \sqrt{1- \frac{4 \sigma A}{B^2}}\right].
\end{equation}
Eq.~\eqref{eq:kpm} can have zero, one, or two positive real solutions. In the latter two cases the band of wavenumbers corresponding to positive real eigenvalues is $\left[0,k_\pm\right]$ and $\left[k_-,k_+\right]$, respectively.\footnote{Provided that $\text{tr}\widetilde{\tens{L}}$ is negative at the roots. For a positive trace the corresponding root belongs to the subdominant eigenvalue (``$-$''sign in Eq.~\eqref{eq:lambda}), and hence the dominant eigenvalue is positive and has no root at $k_\pm$.}
If $B<0$ only $k_-$ can be real and only if $A=f_u g_v-f_v g_u<0$. The onset occurs at $k_-=0$ for $A=0$, i.e.,
\begin{equation}\label{eq:CH}
f_u = g_v^{-1}f_v g_u\,.
\end{equation} 
This corresponds to an Allen-Cahn instability (Table~\ref{tab:instab}).
If $B>0$, both $k_+$ and $k_-$ are real if $\frac{B^2}{4 \sigma}\geq A\geq 0$. A Turing instability occurs if $k_+=k_-$, i.e., with critical wavenumber
\begin{equation}\label{eq:kT}
k_T^2 =k_\pm^2=\frac{B}{2 \sigma}=\frac{\sigma f_u + g_v }{2 \sigma}= \frac{g_v \pm \sqrt{-\sigma f_v g_u}}{\sigma}\,.
\end{equation} 
This instability appears at $A=\frac{B^2}{4 \sigma}$, i.e., at $\sigma f_u = g_v  \pm 2 \sqrt{-\sigma f_v g_u}$ if the trace
\begin{equation}\label{eq:trace}
\tr \widetilde{\tens L} = f_u + g_v -  k_T^2 (1+ \sigma)
\end{equation}
 is negative, otherwise it would correspond to a minimum of the dispersion relation instead of a maximum.
That is, for an RD system a Turing instability requires at least one species to be autocatalytic ($B>0$) and, additionally, $(1-  \sigma) (\sigma f_u - g_v)<0$, which is proven by inserting $k_T$~\eqref{eq:kT} into the trace~\eqref{eq:trace} that has to be negative.  Here we choose $u$ as the only autocatalytic species, i.e., we assume $f_u>0, \, g_v<0$, so that a Turing instability only occurs\footnote{In the RD setting, it corresponds to the known requirement that the inhibitor diffuses faster than the activator. That is, changing the roles of $u$ and $v$ alters the condition to $\sigma<1$. If both, $f_u$ and $g_v$, are either positive or negative a Turing instability cannot occur.} if $  \sigma >1 $ at
\begin{equation}\label{eq:Turing}
\sigma f_u = g_v  + 2 \sqrt{-\sigma f_v g_u}> |g_v|\,.
\end{equation} 
This means that nonreciprocity via $f_v g_u<0$ is another necessary condition, i.e., a reciprocal interaction always prevents a Turing instability.
When $A$ crosses zero for $B>0$, $k_-$ becomes again complex, i.e., then $k_+$ is the only remaining root of the dispersion relation. However, this does not correspond to an Allen-Cahn instability, instead the Turing band of unstable wavenumbers simply attaches to $k=0$ when $A\leq 0$.

The loci of all oscillatory instabilities, i.e., at onset with $\Re \tilde\lambda = 0$ and frequency $\Im \tilde\lambda = \tilde\omega \ne 0$, are determined by $\tr \widetilde{\tens{L}}=0$, i.e., if $f_u +   g_v - k^2 (1+   \sigma)=0$. This gives marginally stable modes with
  \begin{equation}\label{eq:ko}
k_\text{o}^2 = \frac{f_u +   g_v }{1+   \sigma},
\end{equation}
and oscillations first occur at $k_\text{o} = 0$, i.e., only a Hopf instability (Table~\ref{tab:instab}) is possible. Its threshold is at 
\begin{equation}\label{eq:Hopf}
f_u +   g_v =0
\end{equation}
 with $\tilde\omega_0=\det\widetilde{\tens{L}}(k=0)=A$.

Next, we compare stationary and oscillatory instabilities by considering the transition from real to complex eigenvalues. Complex eigenvalues $\widetilde \lambda$ occur for $(\tr{\tens{\widetilde L}})^2 - 4 \det{\tens{\widetilde L }}<0$, i.e., if 
\begin{equation}
  [f_u -   g_v - \left(1-   \sigma \right) k^2]^2 + 4 f_v g_u < 0\,.
  \label{eq:real-oscil-k}
\end{equation}
Similar to the Turing instability, any oscillatory instability requires nonreciprocal interactions $f_v g_u<0$. 
In consequence, for $k=0$ the condition~\eqref{eq:real-oscil-k} can be reduced to
\begin{equation}
A> \frac{1}{4 } (f_u +   g_v)^2
\end{equation}
which is outside the parametric region where a large-scale stationary  instability is observed ($A<0$) implying that our previous consideration on the Allen-Cahn instability applies.

For $k=k_T$ we introduce \eqref{eq:kT} into~\eqref{eq:real-oscil-k} and find that the eigenvalue is complex if
\begin{equation}
 g_v - \frac{4 \sigma}{1+ \sigma  } \sqrt{- f_v g_u} < \sigma f_u<  g_v + \frac{4 \sigma}{1+ \sigma  } \sqrt{- f_v g_u} \,.
\end{equation}
The onset of the Turing instability [cf.~\eqref{eq:Turing}] is outside of this interval if $\sigma   \ne 1$. Then our previous considerations based on real eigenvalues apply to the Turing instability as well. For the special case $\sigma=1$, complex eigenvalues occur independently of the wavenumber at $\sigma f_u = g_v  \pm 2 \sqrt{-\sigma f_v g_u}$. This is identical to the onset condition of the Turing instability [cf.~\eqref{eq:Turing}]. Furthermore, at this specific point one has $k_T=k_\text{o}$. Thus, the Turing instability is prohibited by complex eigenvalues if $\sigma=1$, as one would expect for a Turing system with equal diffusion constants.

\begin{figure}[t]
\centering
\includegraphics[width=0.8\textwidth]{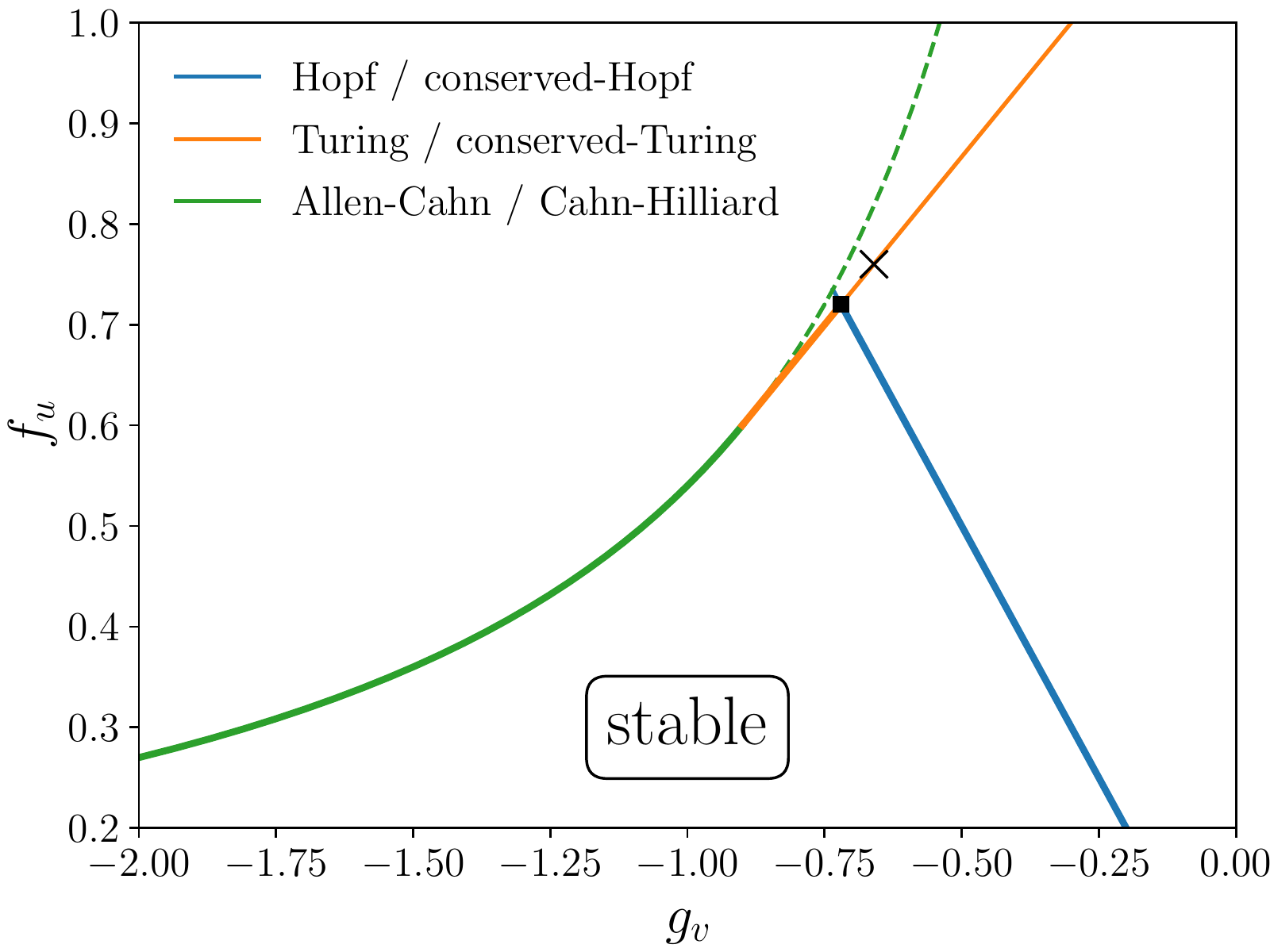}~
\caption{Linear stability diagrams in the $(g_v,f_u)$-plane showing the thresholds of Hopf/conserved-Hopf [homogeneous/large-scale oscillatory, Eq.~\eqref{eq:Hopf}], Turing/conserved-Turing [small-scale stationary, Eq.~\eqref{eq:Turing}], and Allen-Cahn/Cahn-Hilliard [homogeneous/large-scale stationary, Eq.~\eqref{eq:CH}] instabilities as blue, orange and green lines, respectively. Here, $\sigma=1.5>1$ and nonreciprocity $f_v g_u=-0.54$. The focus lies in the region where $u$ is autocatalytic ($f_u>0$) and $v$ is not ($g_v<0$). The linearly stable region is delimited by heavy solid lines. Thin solid lines indicate where further instabilities set on beyond the dominating one. The dashed green line indicates where the already unstable Turing band reaches $k_-=0$ (transition across orange line) or where the unstable complex eigenvalues near $k=0$ become real (transition across blue line). The square symbol marks the codimension-two point [Eq.~\eqref{eq:a22cd2}] where Hopf/conserved-Hopf and Turing/conserved-Turing instabilities occur simultaneously. The cross symbol indicates the loci of the dispersion relation given in Fig.~\ref{fig:dispersion-relations}.
} 
\label{fig:stab-diagram} 
\end{figure}

 A codimension-two point exists if Hopf ($f_u +   g_v=0$) and Turing ($\sigma f_u = g_v + 2 \sqrt{-\sigma f_v g_u}$) instability occur simultaneously, i.e., if
\begin{equation}
g_v = -f_u = -2 \frac{\sqrt{-\sigma f_v  g_u }}{1+  \sigma}
\label{eq:a22cd2}
\end{equation}
and all aforementioned requirements are fulfilled, too.
A typical stability diagram in the ($g_v,f_u$)-plane at fixed  $f_v$, $g_u$ and $\sigma$ is given in Fig.~\ref{fig:stab-diagram}. 

Although onset conditions for linear instabilities for the two-species RD system and the corresponding nonreciprocal two-field CH model are identical, the respective dispersion relations are not. In particular, for the large-scale oscillatory instability in the conserved case the frequency scales with $k^2$, i.e., $\omega_0=k^2 \tilde \omega_0$. Therefore, directly at onset the large-scale instability cannot be oscillatory as there $k_\text{o}=0$, and the conserved-Hopf instability differs from the standard Hopf  instability at $f_u=-g_v$, and takes place only when a mode with the largest available wavelength becomes unstable. For a one-dimensional finite sized system with domain length $L$ and periodic boundary conditions the available wavenumbers are $k_{L/n}= 2n\pi/L$.

\begin{figure}
\centering
\includegraphics[width=0.8\textwidth]{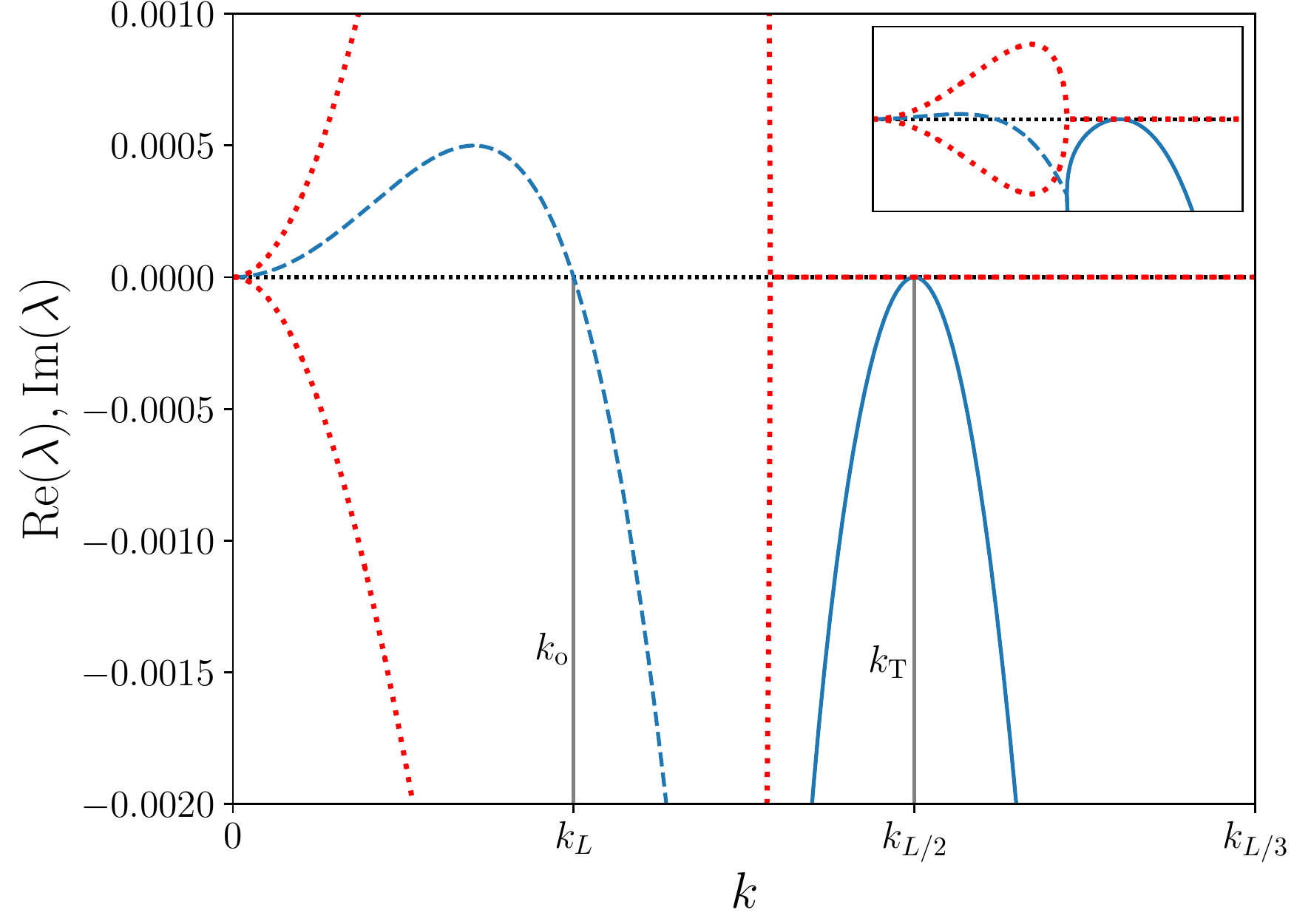}
\caption{(a) Dispersion relation showing real and imaginary part of $\lambda(k)$  [Eq.~\eqref{eq:lambda}] at the parameter values where the conserved-Turing instability has its onset (at $k_{L/2}=k_T$) and is resonant with the critical mode of the conserved-Hopf instability ($k_o=k_L$), i.e., $k_T= 2k_o$. Solid and dashed blue lines represent $\mathrm{Re}\,\lambda(k)$ for real and complex eigenvalues, respectively. Dotted red lines give the imaginary parts. The inset illustrates the dispersion on a larger magnitude range.  Parameters are $\sigma=1.5$, $f_u=0.76$, $f_v=-1$, $g_u=0.54$ and $g_v=-0.66$ and correspond to the cross symbol in Fig.~\ref{fig:stab-diagram}. In a finite system this resonance is only realizable for a perfectly tuned domain size of $L=10 \pi$.}
\label{fig:dispersion-relations} 
\end{figure}

As a consequence of this effect, resonances only occur in the vicinity of but not directly at the codimension-two point of the infinite system. In the most interesting case, an oscillatory mode (wave) with $k_L =k_\text{o}$
and a stationary  mode with the wavenumber $k_{L/n}=k_T$ where $n>1$ are simultaneously marginal. Fig.~\ref{fig:dispersion-relations} presents a corresponding dispersion relation  with $n=2$, i.e., for the parameters marked by a cross symbol in Fig.~\ref{fig:stab-diagram}. If one considers larger values of $n$ the position of the cross moves on the orange line closer to the codimension-two point. For specific finite systems of domain size $L\neq2\pi n/k_T$, the relevant stationary modes are rather related to $k_-$ or $k_+$ -- the limiting values of the band of unstable conserved-Turing modes [cf.~Eq.~\eqref{eq:kpm}].
Close to the corresponding primary bifurcations, the behavior can be analyzed with the help of a weakly nonlinear analysis \cite{PiRu1999csf,YDZE2002jcp}, as described in section~\ref{sec:resonance} for a general system \eqref{91uwas}. Further beyond the onset, one may compare general weakly nonlinear results with fully nonlinear time simulations for a specific conserved amended FitzHugh-Nagumo model (see  section~\ref{sec:resonance-specific}). Note that for systems without conservation laws resonances are frequently studied. Examples include  Hopf-Turing, Turing-Turing, and wave-Turing resonances in reaction-diffusion systems or nonlinear optical systems \cite{DLDB1996pre,PiRu1999csf,DeW1999acp,YDZE2002prl,YDZE2002jcp}. 

\section{Hopf--Turing resonance - weakly nonlinear analysis}
\label{sec:resonance}

We consider the resonant interaction between a conserved-Turing and two conserved-Hopf modes with the three wave vectors forming an isosceles triangle. In a finite system, they satisfy the condition $\vec{k}_{o1}-\vec{k}_{o2} + \vec{k}_\pm=0$ with $|\vec{k}_{o1}|=|\vec{k}_{o2}|\equiv k_o$. 
The corresponding interactions for a nonconserved system have been analyzed in Ref.~\cite{PiRu1999csf}. We consider the case when the resonance occurs close to the common onset of linear instability for a specific finite system, defined by the critical values of two parameters. We impose a small deviation in one of these parameters, let us say $\beta=\beta_c + \varepsilon$ with $|\varepsilon|\ll 1$. Then, both the growth rates and the amplitudes are small and, thus, by expanding them in powers of $\varepsilon$, they are treated through a weakly nonlinear approach. Details are given in Appendix~\ref{app}, here we only sketch the procedure and give its results.

As an ansatz, we write the two-component vector field $\vec u = (u,v)$ as a sum of the uniform steady state $\vecg u_s$ and a small deviation, i.e.,
\begin{equation}\label{eq:ansatz}
\vec u = \vec u_s + \varepsilon \left[a_+(T) \vec u_+ e^{\text{i} \vec k_+ \cdot \vec x} +a_{o1}(T) \vec u_{o1} e^{\text{i} \left(\vec k_{o1} \cdot \vec x + \omega_{o1} t\right)} +a_{o2}(T) \vec u_{o2} e^{\text{i} \left(\vec k_{o2} \cdot \vec x + \omega_{o2} t\right)} + \text{c.c.}\right] + \mathcal{O}(\varepsilon^2)
\end{equation}
where $a_+(T)$, $a_{o1}(T)$, $a_{o2}(T)$ are the amplitudes that evolve on a large timescale $T=\varepsilon t$; $\vec u_+ \in \mathbb{R}$ and $\vec u_{o1}, \vec u_{o2}\in \mathbb{C}$ are the zero eigenvectors of the stationary and the two wave modes, respectively. Analogously to the zero eigenvalues they are equal in the conserved and nonconserved case. The frequencies $\omega_{o1}$ and $\omega_{o2}$ are the imaginary parts of the eigenvalues at onset of instability of the wave modes. We consider an isotropic system, which implies that $\vec u_{o1} = \vec u_{o2} \equiv \vec{u}_o$ and $\omega_{o1}=\omega_{o2} \equiv \omega_o$. Note that either end of the Turing-unstable wavenumber band, $\vec{k}_+$ or $\vec{k}_-$, may be used for the stationary mode. Here, we take $\vec{k}_+$.

After inserting Eq.~\eqref{eq:ansatz} into Eqs.~\eqref{91uwas}, the leading-order amplitude equations are obtained at $\mathcal{O}(\varepsilon^2)$ by applying solvability conditions, i.e., multiplying by corresponding adjoint eigenvectors $\vec{u}_+^\dagger$, $\vec{u}_o^\dagger$ normalized to satisfy $\vec{u}_+^\dagger \cdot \vec{u}_+= \vec{u}_o^\dagger \cdot \vec{u}_o=1$, and projecting onto the extant Fourier modes. For details see Appendix~\ref{app-a}.

The general form of the resulting lowest-order resonant amplitude equations is the same as in the standard nonconserved case. However, the coefficients of these equations, that depend on the eigenvectors, the Jacobian matrix and the Hessian, carry an additional $k_+^2$ and $k_o^2$ prefactor corresponding to the respective stationary and oscillatory mode:
 \begin{align} 
 \begin{split}\label{9ampl}
 \dot a_+ &= k_+^2 (\mu_+ a_+ + \nu_+ \bar a_{o1} a_{o2} )~\\
\dot a_{o1} & = k_o^2(\mu_o a_{o1} + \nu_o \bar a_{+} a_{o2} ), \quad 
\dot a_{o2}  =  k_o^2(\mu_o a_{o2} + \nu_o a_+  a_{o1})
\end{split}
\end{align}
where $\nu_+$
is real, while  $\nu_o $ 
is complex. The coefficient $\mu_+$ is real and, since the imaginary part of $ \mu_o$ can be absorbed into the frequency of the wave modes (i.e., by applying a transformation $a_{o1/2} \to a_{o1/2}e^{i k_o^2 \text{Im}\mu_o t}$ or equivalently $\omega_o \to \omega_o + k_o^2 \text{Im} \mu_o$), this parameter can be also viewed as real. Since the interaction coefficients are generally distinct, the system lacks gradient structure, allowing, in principle, for persistent nonstationary behavior within the amplitude-equation representation, leading to secondary oscillations on an extended scale. Including cubic terms is unnecessary close to the onset, since the amplitudes may remain finite in this system even without higher-order damping interactions.

\begin{figure}
\centering
   \includegraphics[width=0.55\hsize]{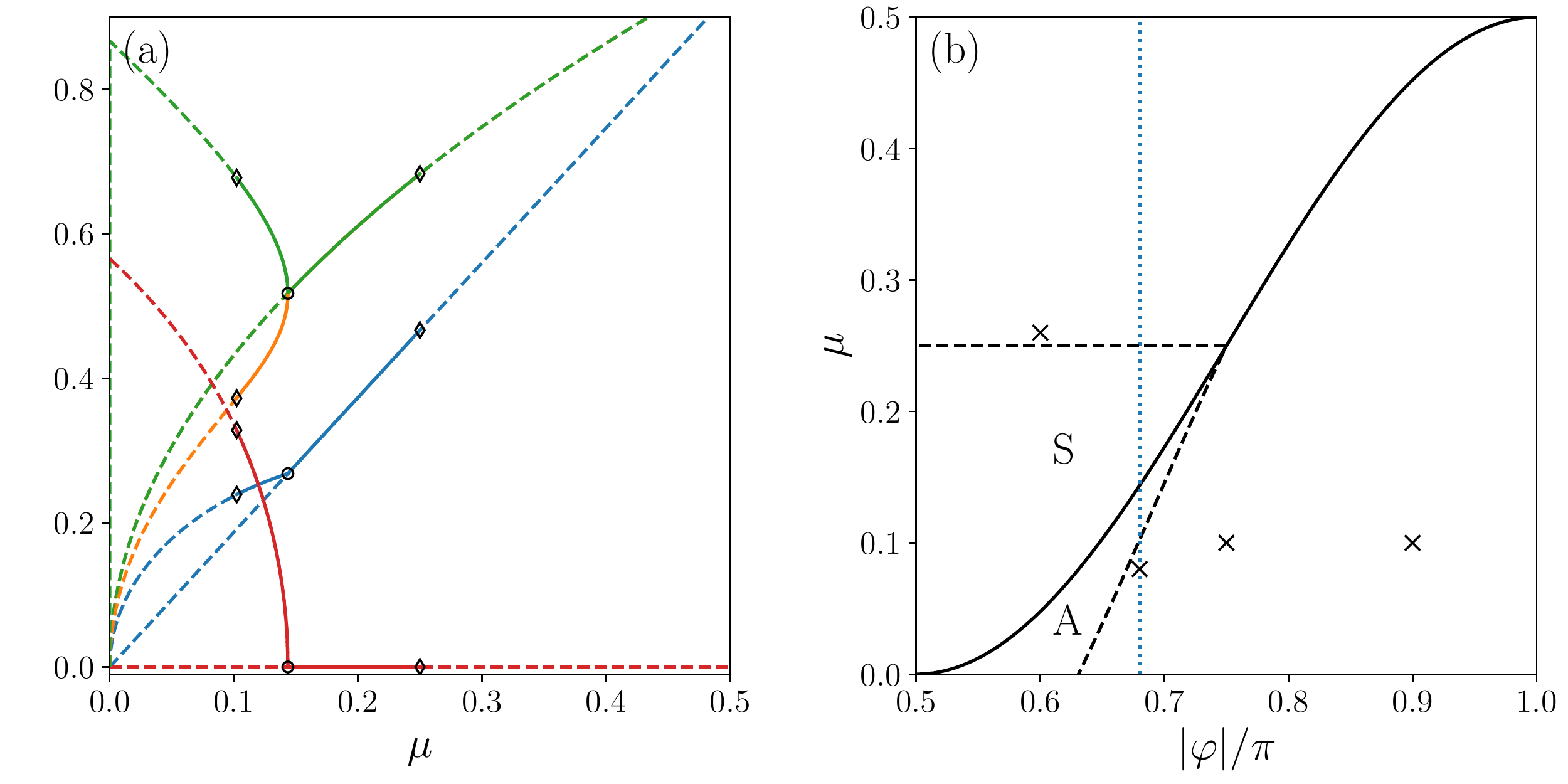}~\\
   \vspace{-0.1cm}
  \includegraphics[width=0.233\hsize]{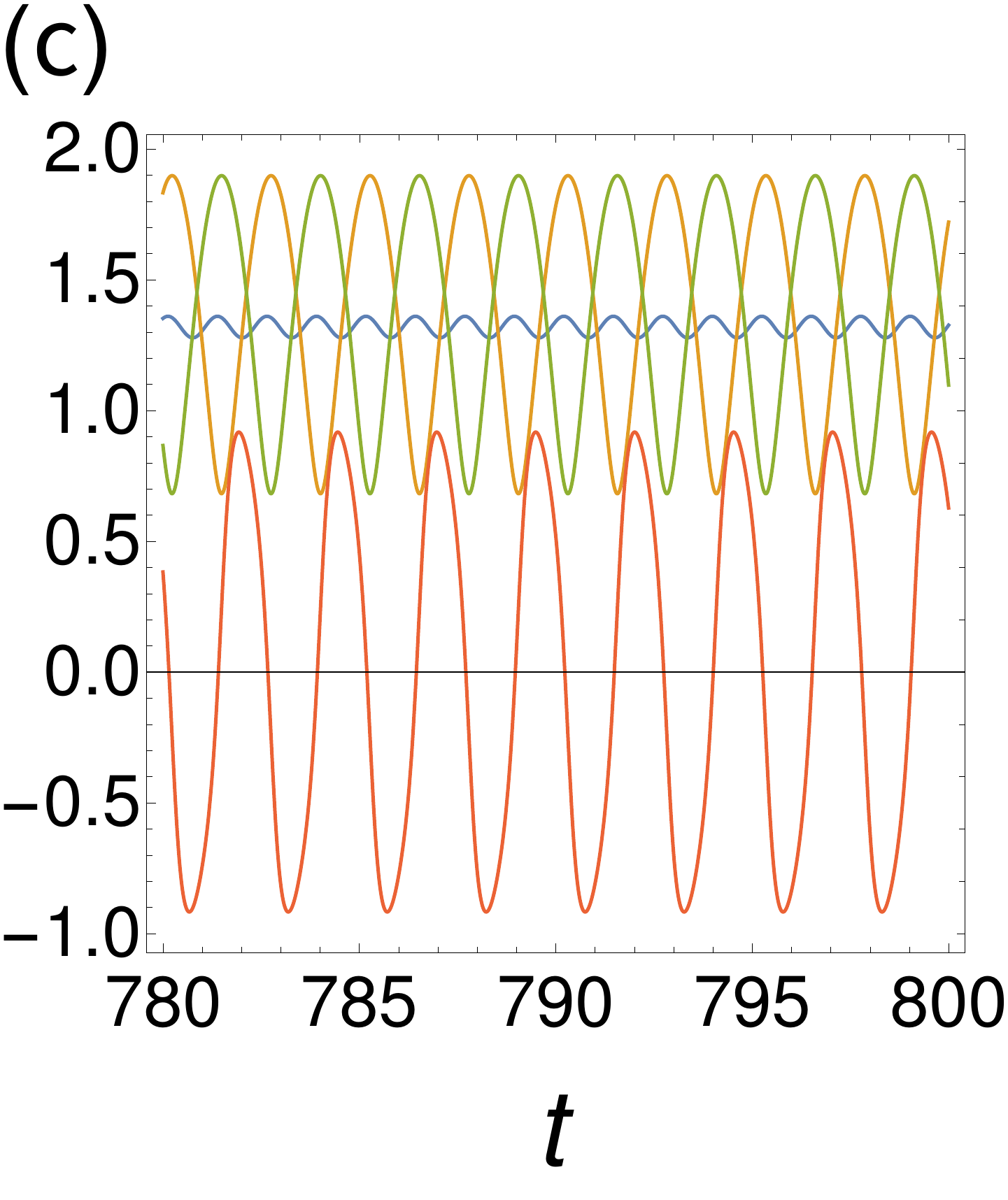}\hspace{0.01\hsize} 
  \includegraphics[width=0.233\hsize]{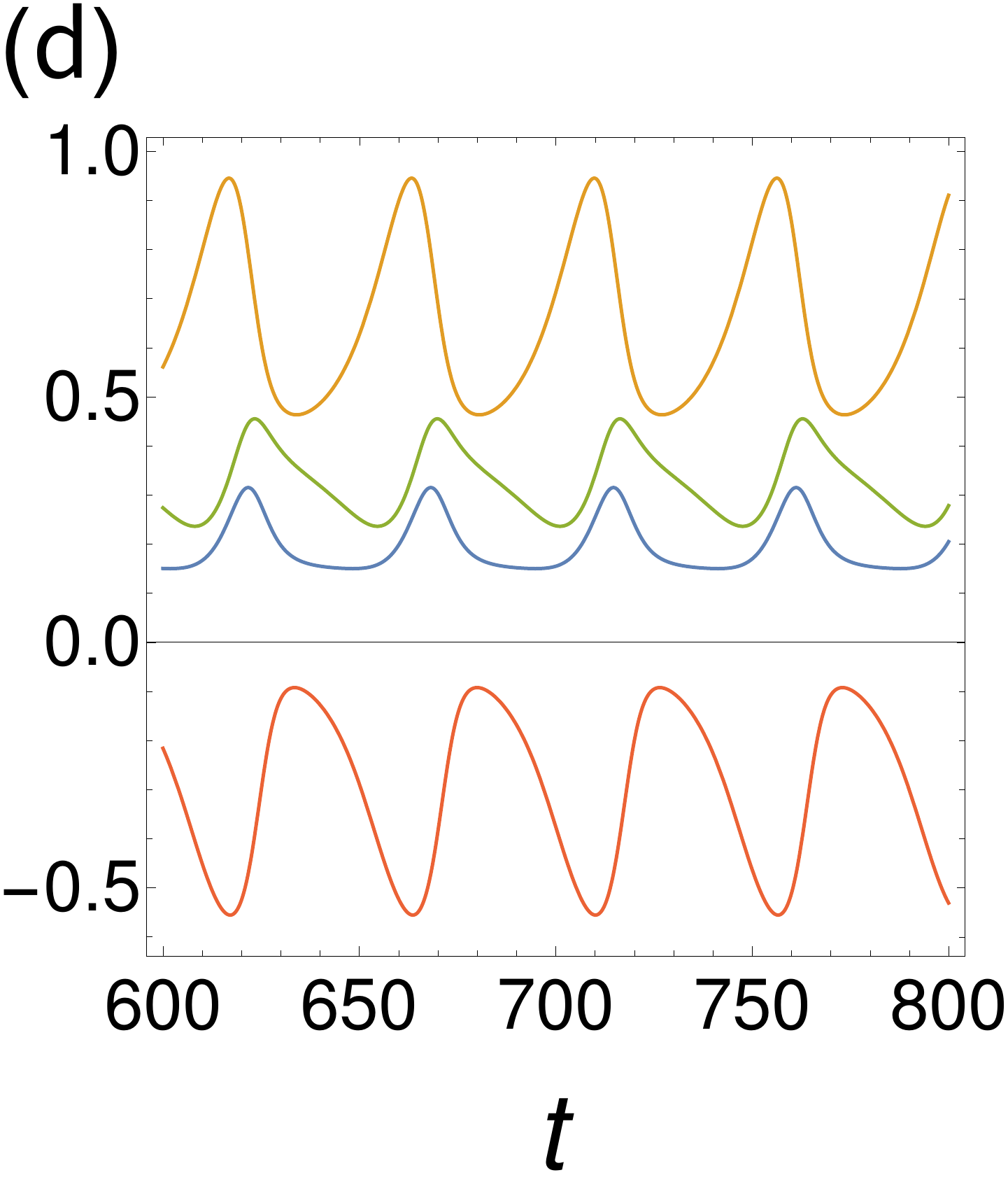}\hspace{0.01\hsize} 
  \includegraphics[width=0.233\hsize]{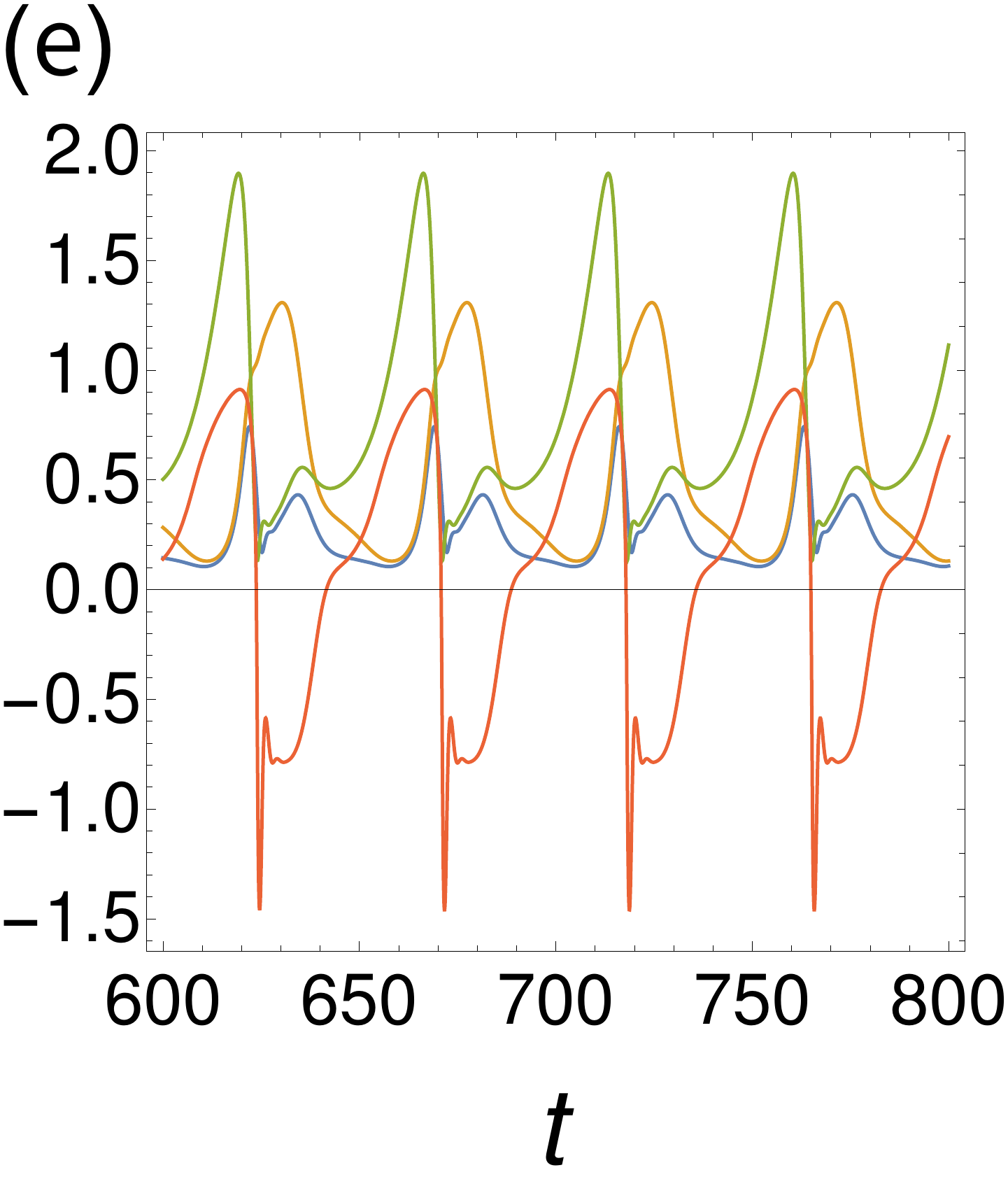}\hspace{0.01\hsize} 
   \includegraphics[width=0.23\hsize]{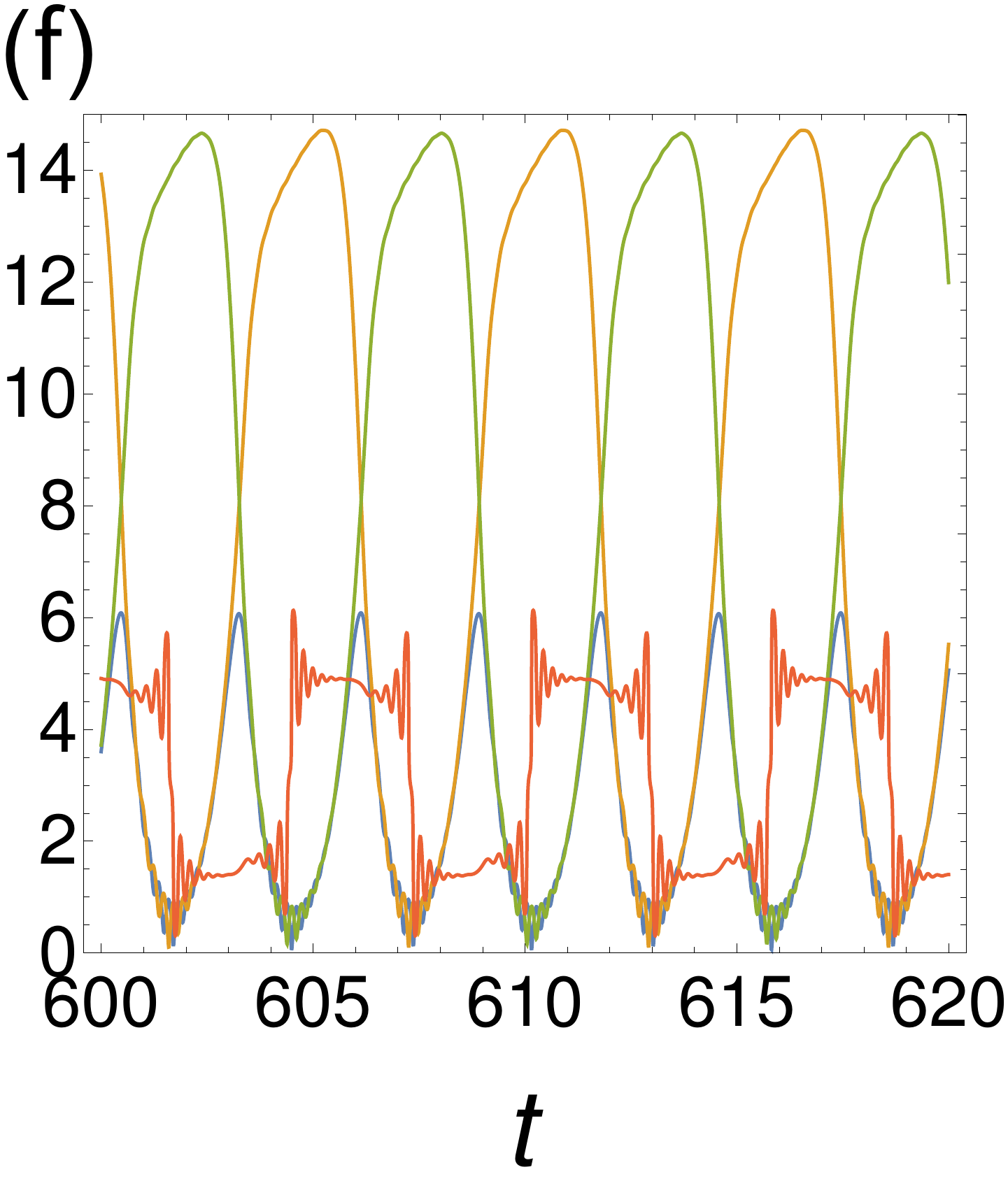}\\
   \vspace{-0.1cm}
  \includegraphics[width=0.25\hsize]{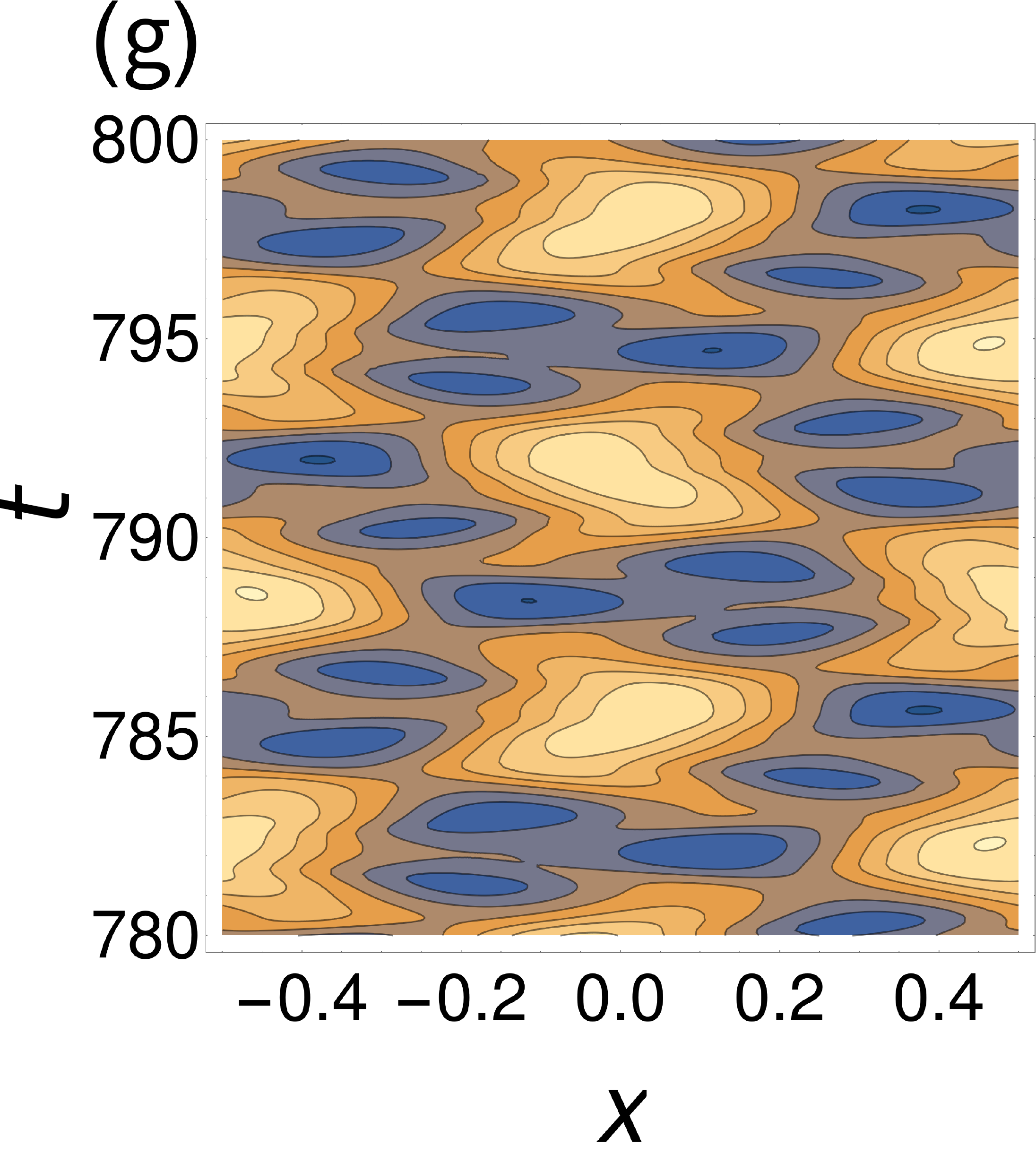}\hspace{0.015\hsize}
  \includegraphics[width=0.23\hsize]{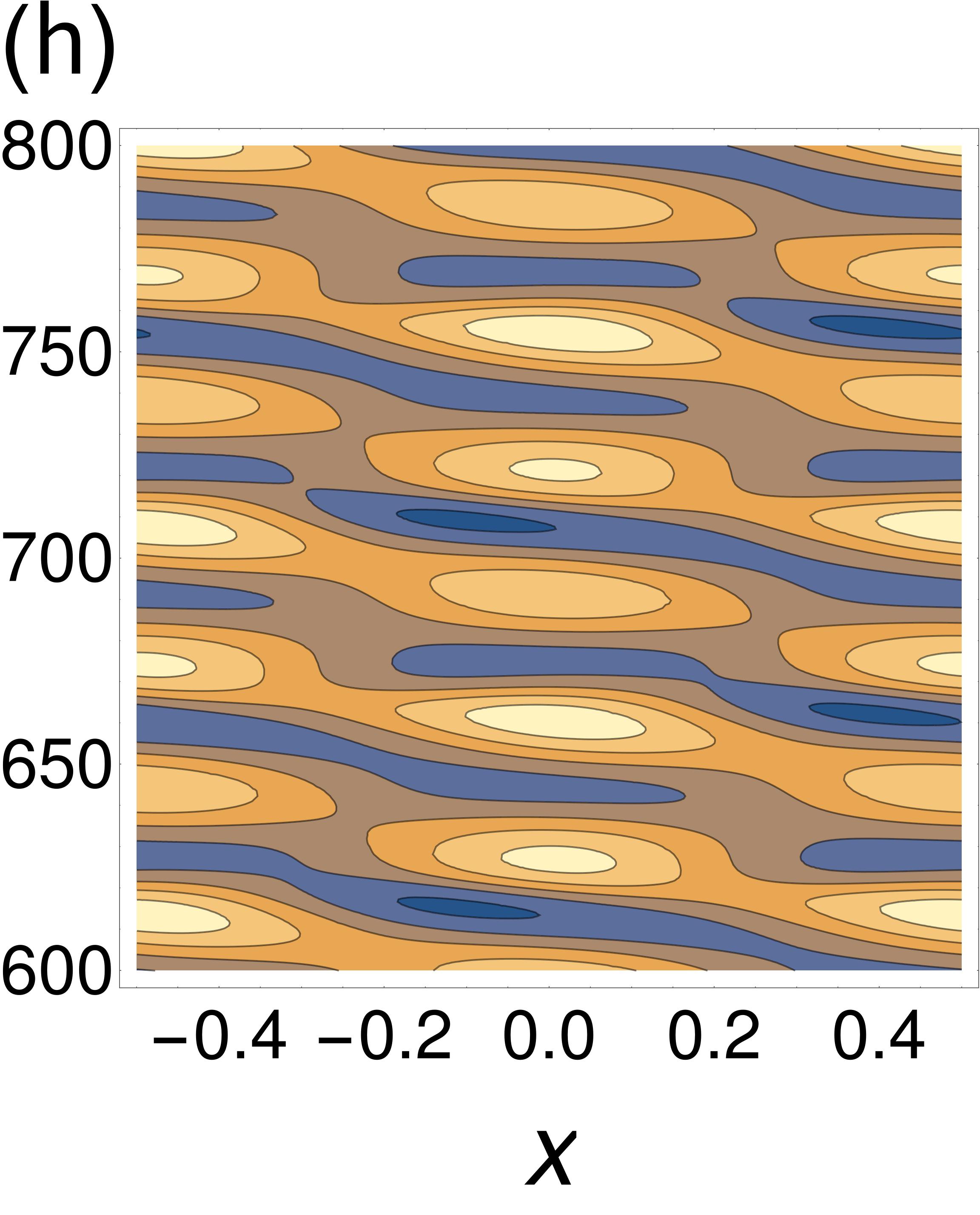}\hfill
  \includegraphics[width=0.23\hsize]{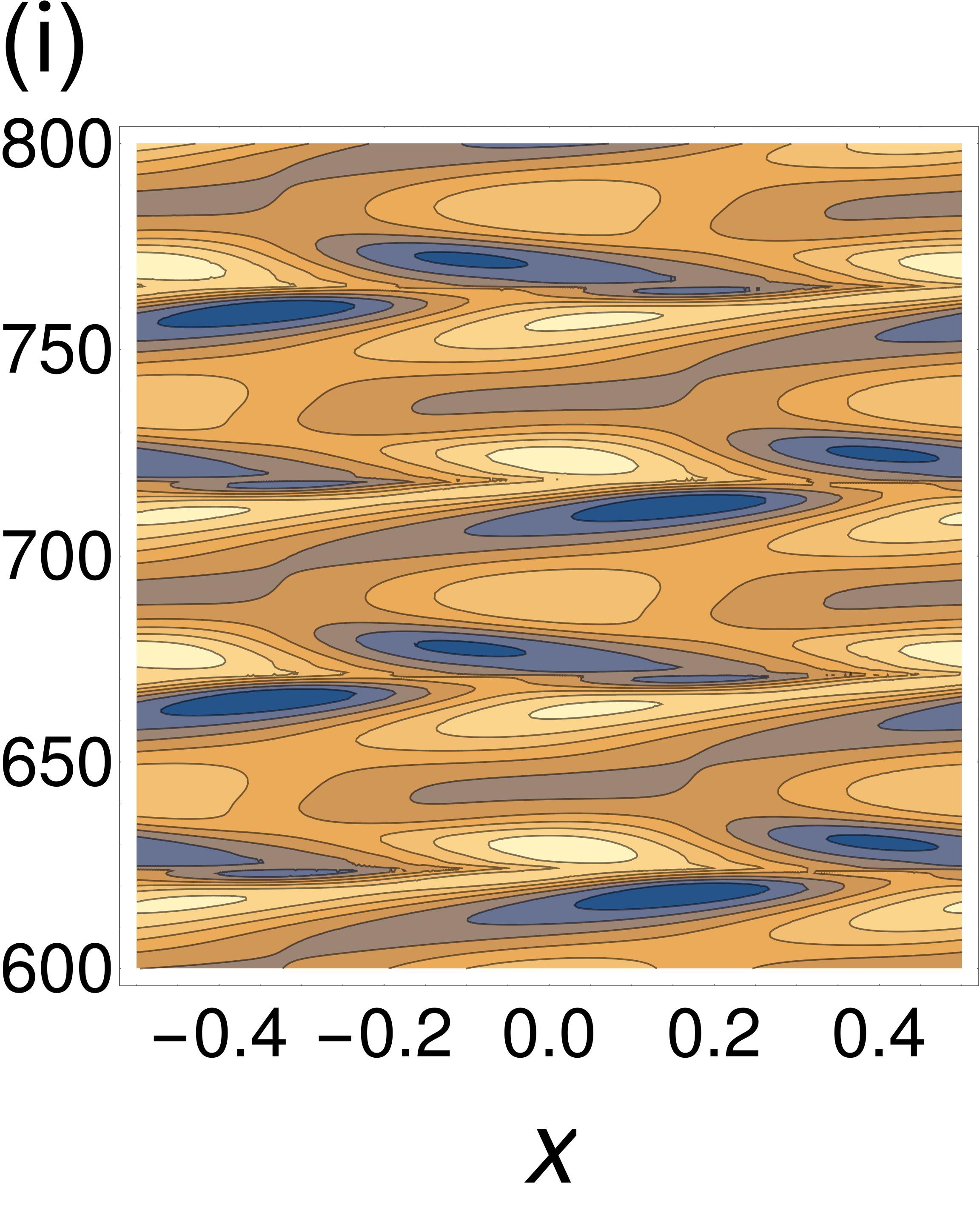}\hfill
   \includegraphics[width=0.23\hsize]{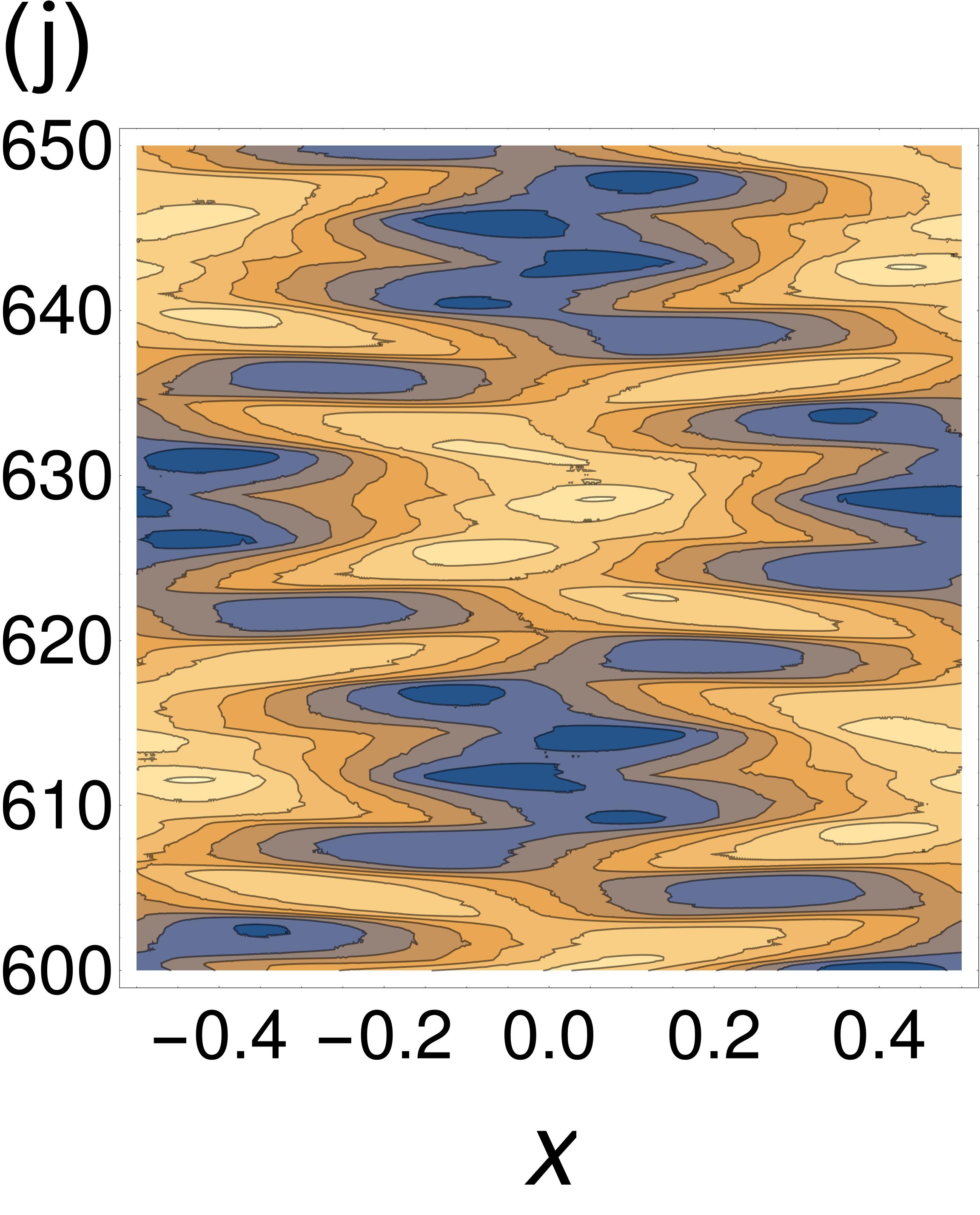}~\\
\vspace{-0.1cm}
     \includegraphics[width=0.25\hsize]{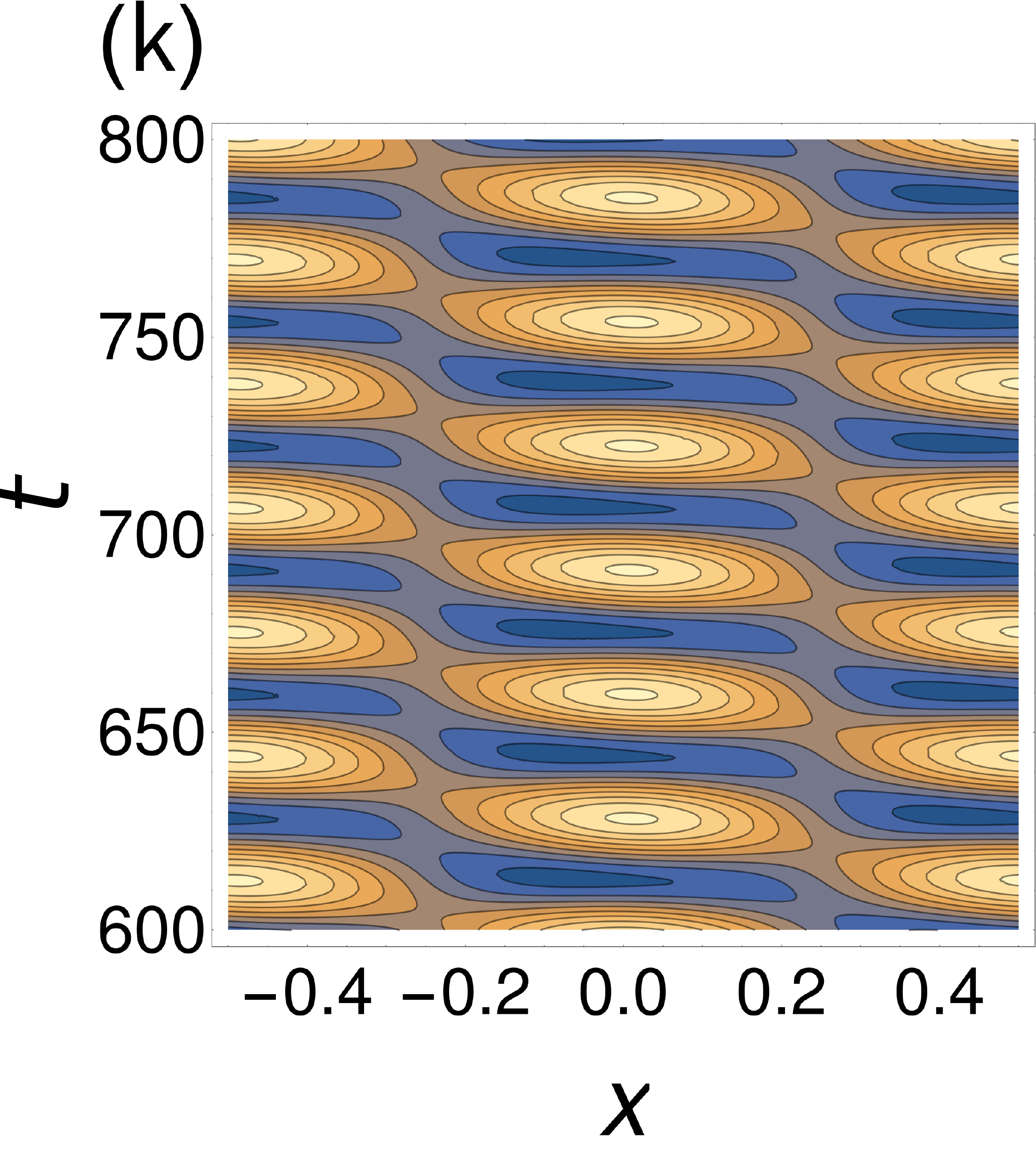}\hspace{0.05\hsize} 
   \includegraphics[width=0.23\hsize]{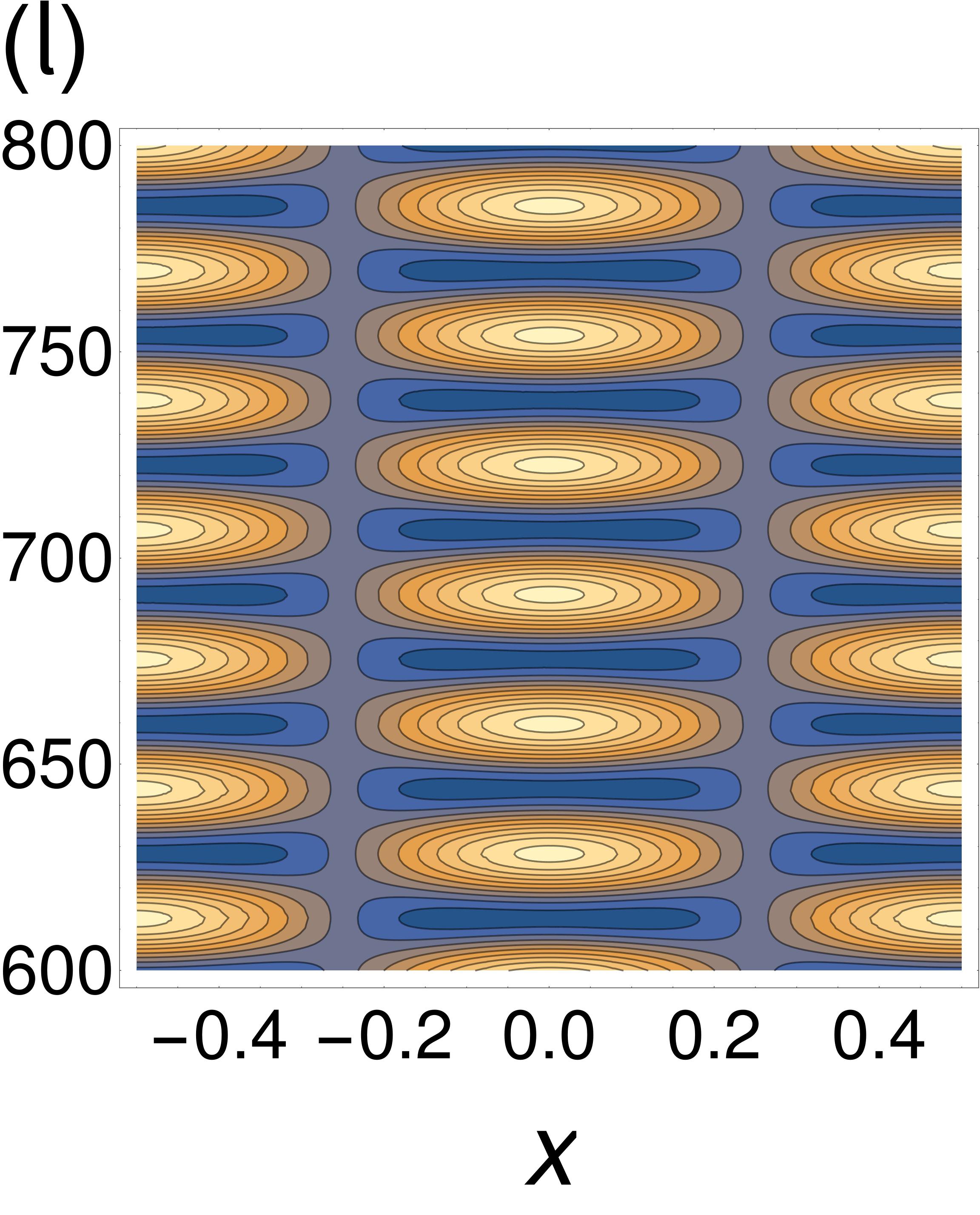}
 \caption{{\footnotesize Results of the weakly nonlinear analysis are presented: (a) Branches of stationary solutions at $ \varphi= 0.68 \pi$ as a function of $\mu$. Diamond and circle symbols indicate Hopf and pitchfork bifurcations. Stable and unstable branches of solutions are shown by the solid and dashed lines, respectively. (b) Phase diagram with solid and dashed lines indicate pitchfork and Hopf bifurcations, respectively. The regions of prevailing stationary linearly stable symmetric and asymmetric states are indicated by the letters ``S'' and ``A'', respectively. (c)-(f) Amplitudes as functions of time  for different types of long-time oscillatory behavior. The parameters are (c) $\mu=0.26, \,\varphi=0.6 \pi$, (d) $\mu=0.08, \,\varphi=0.68 \pi$, (e) $\mu=0.1, \,\varphi=0.75 \pi$ and (f) $\mu=0.1, \,\varphi=0.9 \pi$, and  are indicated by cross symbols in the panel~(b). In (a,c-f) the curves for $\rho_+,  \rho_1, \rho_2$, and $\Theta$ are shown as blue, orange, green, and red lines, respectively. Panels~(g)-(l) give space-time plots in the specific 1d setup of the original field given by $u(x,t)=\rho_+ \cos(k_+ x + \Theta(t)) + \rho_1 \cos(k_o x + \omega t) + \rho_2 \cos(-k_o x + \omega t)$ with $k_+=2 k_o = 4\pi/L$ with domain size $L=1$ and (g) $\omega=1$, (h)-(l) $\omega=0.2$. Panels~(g) to (j) correspond to the oscillating states of the respective panels (c) to (f). Panels~(k) and (l) correspond to the linearly stable asymmetric state at $\mu=0.12,\,\varphi=0.68\pi$ and the linearly stable symmetric state at $\mu=0.2,\,\varphi=0.68\pi$, respectively.}}  
  \label{fig:f1wt} \end{figure}

Using the polar representation of the complex amplitudes, $a_+ = \rho_+\; \E^{\I \theta_+}, \; a_{o1} = \rho_1 \E^{\I \theta_1}, \; a_{o2} = \rho_2 \E^{\I \theta_2}$, (\ref{9ampl}) is written in terms of the three real positive amplitudes and three phases, but dynamics depends only on the single phase combination $\Theta = \theta_+ + \theta_1 - \theta_2$, so that (\ref{9ampl}) can be reduced to a system of four real equations as described in Appendix~\ref{app-a}:
\begin{align} 
\begin{split}\label{1amplreal} 
\dot{\rho}_+ & =  -\rho_+ + \rho_1 \rho_2 \cos \Theta\,, \\
\dot{\rho}_1 & =  \mu \rho_1 +  
     \rho_+ \rho_2 \,\cos (\Theta - \varphi)\,, \quad 
\dot{\rho}_2 = \mu \rho_2 +  
     \rho_+ \rho_1 \,\cos (\Theta + \varphi)\,, 
     \end{split}   \\ 
\dot{\Theta} & = - \frac{\rho_1 \rho_2}{\rho_+} \sin \Theta
  -  \rho_+ \left[ \frac{\rho_1 }{\rho_2} \sin (\Theta + \varphi) + 
     \frac{\rho_2}{\rho_1} \sin (\Theta - \varphi) \right]. \label{1amplim}
\end{align} 
This system of equations has stationary and oscillatory solutions summarized in Fig.~\ref{fig:f1wt}.
In the stationary case the values of the amplitudes $\rho_+,\rho_{1,2}$ obtained by resolving (\ref{1amplreal}) are (see Appendix~\ref{app-b} for details)
\begin{equation}
{\rho}_+ = \frac{|\mu|}{[\cos(\Theta-\varphi)\cos(\Theta+\varphi)]^{1/2}}, \quad
{\rho}_{1,2} = \left[-\frac{\mu}{\cos(\Theta \pm\varphi)\cos \Theta}\right]^{1/2}. 
\label{1statrho}    \end{equation}

Introducing (\ref{1statrho}) into (\ref{1amplim}) brings the equation defining stationary values of $\Theta$ to the form 
\begin{equation} 
- \tan \Theta+\mu[ \tan (\Theta-\varphi)+ \tan (\Theta+\varphi)]= 0.
      \label{1eqth}  \end{equation} 

    A trivial solution to \eqref{1eqth} is $\Theta =0$ which defines the symmetric stationary solution of \eqref{1amplreal} with $\rho_+ = \rho_1^2 = \rho_2^2 = |\mu/\cos \varphi|$. This state corresponds in the original model to a steady pattern with a superposed standing wave giving the impression of a mass oscillating between two neighboring peaks (for a visual impression in the 1d setup see Fig.~\ref{fig:f1wt}~(l)). 
       Related localized patterns with a superposed oscillation on a larger scale are also found in a nonreciprocal Cahn-Hilliard model \cite{FrTh2021ijam}. In active phase-field-crystal models, related states are described as alternating localized or alternating periodic states \cite{VHKW2022msmse}.

An asymmetric stationary solution is obtained when \eqref{1eqth} is converted through a chain of trigonometric transformations (see  Appendix~\ref{app-b}) into a transparent implicit relation
\begin{equation} 
 (1-2\mu)\cos^2 \Theta= \sin^2\varphi.
      \label{1eqtha}  \end{equation} 
 Hence, the asymmetric solution is confined to the interval $0\leq \mu\leq\frac 12 \cos^2\varphi$. The existence limits correspond to the bifurcation from the trivial state at $\mu=0$ and the pitchfork bifurcation from the symmetric solution at $\mu=\frac 12 \cos^2\varphi$, respectively. An additional restriction comes from the requirement for the amplitudes given by \eqref{1statrho} to be real and positive. This requires $|\varphi|>\pi/2$. Branches of solutions do not diverge if $|\Theta|<\varphi-\pi/2$. 
 
The asymmetric state corresponds in the original model to a pattern with a superposed traveling wave of half the wavenumber giving the impression of mass unidirectional traveling between peaks (see Fig.~\ref{fig:f1wt}~(k)). The asymmetric state is stable beyond the pitchfork bifurcation of the symmetric solution, and undergoes a Hopf bifurcation on another stability limit. Note that this is a secondary bifurcation on top of the Hopf bifurcation creating the linearly growing wave modes involved in this planform.  In the original model this Hopf bifurcation results in a state corresponding to a standing wave with a superposed traveling modulated wave (see Figs.~\ref{fig:f1wt}~(h)-(j)). To our knowledge such states have not yet been systematically studied in systems with conservation laws although some states of similar complexity are described for an active phase-field-crystal model for a mixture of active and passive particles \cite{VHKW2022msmse}.
The instability limits of both symmetric and asymmetric stationary solutions merge at the double zero singularity located at $\varphi= \frac34 \pi, \mu=\frac14$.  Further details, involving elaborate calculations, are given in the Appendix~\ref{app-c}.
The pitchfork bifurcation corresponds in the original model to a kind of a drift-pitchfork bifurcation. Related localized patterns with superposed oscillation and drift are also found in a nonreciprocal Cahn-Hilliard model  \cite{FrTh2021ijam} and in active phase-field-crystal models \cite{OKGT2020c,VHKW2022msmse}. Note that, due to the resonance, the described complex scenario differs from the basic codimension-1 scenario where a steady state starts to drift at a drift-pitchfork \cite{KrMi1994prl} or drift-transcritical \cite{OpGT2018pre} bifurcation (all called traveling bifurcation in Ref.~\cite{Pismen2006}). It is more closely related to scenarios where a stable standing wave that emerged in a Hopf bifurcation gives way to a modulated traveling wave through a drift-pitchfork bifurcation or where a stationary state that is unstable to a drift mode undergoes an additional Hopf bifurcation \cite{FaDT1991jpi,NiTU2003c}. These scenarios are slightly simpler than the one treated here as they do not involve a spatial resonance. A small deviation from the resonance conditions could also result in such scenarios. This is, however, not captured by the amplitude equations.
  
Fig.~\ref{fig:f1wt}~(a) presents a bifurcation diagram showing the stationary solution branches given by ~\eqref{1statrho} for fixed $\varphi=0.68 \pi$ where the colored lines give the three different amplitudes and the phase as described in the caption. A linear stability analysis of the symmetric state gives $\left[\frac12 \cos^2 \varphi, \frac14\right]$ as the $\mu$-range of linear stability limited by the aforementioned pitchfork bifurcation on the left (circle symbols) and a Hopf bifurcation (diamond symbols) on the right hand border. The latter represents another secondary Hopf bifurcation and for a visual impression of the resulting oscillatory state in the 1d setup see Fig.~\ref{fig:f1wt}~(g).

The discussed existence and linear stability in the $(\varphi,\mu)$-plane are summarized in Fig.~\ref{fig:f1wt}~(b). The pitchfork bifurcation indicated by the circle symbols in panel~(a) is given as the solid line that separates the regions ``A'' and ``S'' where the asymmetric and symmetric stationary solutions are linearly stable, respectively. The stability regions are further limited by the dashed lines that mark the loci of the Hopf bifurcations given by diamond symbols in (a). Examples of corresponding periodic orbits obtained in the description of the amplitude equations are shown in Figs.~\ref{fig:f1wt}~(c) to~(f) in the sequence of increasing $\varphi$; their loci in the $(\varphi,\mu)$-plane are marked by crosses in Fig.~\ref{fig:f1wt}~(b). We have to be warned that the existence region of oscillatory solutions does not encompass the entire domain where stationary solutions are unstable, since, in the absence of cubic and higher-order dumping that is apt to stabilize oscillations in an underlying system, and could be detected by a higher-order bifurcation analysis, some trajectories escape to infinity.
\section{Example: modified FitzHugh--Nagumo system}\label{sec:FHN}
\label{sec:resonance-specific}
Next, we aim at identifying resonant behavior in the fully nonlinear regime. To do so we have to focus on a specific nonreciprocal CH system. We employ for this purpose a simple representative example obtained by choosing $f$ and $g$ in Eqs.~\eqref{91uwas} to be of the third order in intraspecies interactions and linear in interspecies interaction. In particular, we use $f(u,v)=u-u^3-v$ and $g(u,v)=\alpha u-\beta v - v^3$. Correspondingly, $\chi(u,v) = -u^2/2+u^4/4+\beta v^2/(2\gamma) + v^4/(4\gamma) +(1-\alpha/\gamma)uv/2$ in \eqref{eq:energy} as well as $\mu_{u}^\mathrm{nv}=(1+\alpha/\gamma)v/2$ and $\mu_{v}^\mathrm{nv}=-(1+ \alpha/\gamma)u/2$. Both species have nonzero mean densities, i.e., $1/L\,\int u~{\rm d}x=u_s $ and $1/L \, \int v~{\rm d}x= v_s$ that act as effective quadratic nonlinearities in $f$ and $g$, respectively.\footnote{In other words, if we introduce the shifted densities $u-u_s$ and $v-v_s$, the resulting nonlinear terms in the shifted densities include  quadratic nonlinearities.} In the absence of the cubic nonlinearity in $g$,  our example represents a fully mass-conserving version of the standard FitzHugh--Nagumo model. It represents a simple example of \eqref{91uwas}, however, as explained below, here we have not detected the secondary Hopf instabilities
 and the corresponding oscillatory behavior discussed in the previous section. 
Therefore, we include the cubic nonlinearity in $g$ and obtain a conserved modified FitzHugh--Nagumo system that is identical to the recently considered nonreciprocal Cahn-Hilliard model.

For the homogeneous state $(u,v)=(u_s,0)$ with $u_s^2<1/\sqrt 3$ we have $f_u=1-3u_s^2,\, f_v= -1, \, g_v=-(\beta +3 v_s^2), \, g_u=\alpha, \, A=\alpha - (1-3 u_s^2) (\beta +3v_s^2), \, B=(1-3u_s^2)\sigma-(\beta + 3v_s^2)$. Imposing $3 u_s^2<1,\,\beta+3v_s^2>0$, we choose $u$ but not $v$ to be autocatalytic ($f_u>0$, \,$f_v<0$). If the coupling is nonreciprocal ($f_v g_u<0$), i.e., for $\alpha>0$, the necessary conditions for the instabilities are
 \begin{align}
 \begin{split}\label{eq:restrictions}
\text{Cahn-Hilliard:~~}&A<0 \Rightarrow \alpha<(1-3u_s^2)(\beta + 3v_s^2)~\, \\
\text{conserved-Turing:~~}&B>0 \, \, \wedge \, \, \sigma>1 \, \, \wedge \, \, \sigma f_u > g_v  + 2 \sqrt{-\sigma f_v g_u} ~\\
& \Rightarrow \sigma(1-3u_s^2-3v_s^2) >\beta>-\sigma(1-3u_s^2-3v_s^2) + 2 \sqrt{\sigma \alpha} \, \, \wedge \, \, \sigma>1 ~\\
\text{and conserved-Hopf:~~}& A>\frac{1}{4} \left(f_u+  g_v\right)^2 \, \, \wedge \, \, f_u + g_v>0 ~\\
&\Rightarrow \beta< \text{min}\left\{-1 +3u_s^2 -3v_s^2 + 2\sqrt{\alpha},1 - 3u_s^2 -3v_s^2\right\} \\
&\, \, \wedge \, \, \alpha> \left(\frac{1-3u_s^2}{2}\right)^2\,.
\end{split}
 \end{align}
The wavenumbers of stationary and oscillatory marginal modes [cf.~\eqref{eq:kpm} and~\eqref{eq:ko}] are then given by
\begin{align}
k^2_\pm = & k_T^2 \left[1 \pm \sqrt{1- \frac{4 \sigma \left(\alpha-(1-3u_s^2)(\beta +3v_s^2) \right)}{\left((1-3u_s^2)\sigma-(\beta +3v_s^2)\right)^2}}\right]~\label{eq:kpm_FHN}\\
\text{and}\qquad k_\text{o}^2 = & \frac{1-3u_s^2-(\beta +3v_s^2) }{1+  \sigma}\,, \label{eq:ko_FHN}
\end{align}
where $k_T^2=\frac{(1-3u_s^2)\sigma-(\beta+3v_s^2)}{2\sigma}$ is the critical wavenumber at the onset of the conserved-Turing instability.

We consider now a scenario where a marginal conserved-Hopf mode ($k_o=k_L$) and a marginal conserved-Turing mode ($k_\pm=k_{L/2}$) are resonant in a one-dimensional domain. For the considered specific
system this is achieved at $4 k_\text{o}^2 = k_+^2 = k_{L/2}^2 $. Using Eqs.~\eqref{eq:kpm_FHN} and \eqref{eq:ko_FHN} gives after simplification
the critical values
\begin{equation}\label{eq:crit_values}
\alpha_c =\left(1 -3 u_s^2-\frac{16 \pi^2}{L^2} \right) \left(1 - 3 u_s^2 - 
   \frac{4 \pi^2}{L^2} (1 - 3 \sigma)\right)\,, ~\qquad \beta_c = 1 -  3 (u_s^2 +v_s^2) - \frac{4 \pi^2 }{L^2}(1 + \sigma)\,,
\end{equation}
that define this codimension-two point. The corresponding frequency of the marginal conserved-Hopf mode is
\begin{equation}
\omega_o=\frac{8 \sqrt{3} \pi ^3 \sqrt{L^2 (\sigma -1) \left(3 u_s^2-1\right)+4 \pi ^2 (4 \sigma -1)}}{L^4}\,.
\end{equation}
To consider the weakly nonlinear regime in the vicinity of the codimension-two point we set $\alpha=\alpha_c$ and $\beta=\beta_c + \varepsilon$.
The resulting relevant nonzero entries of $\partial \tens{L}(k^2)/\partial \beta$ and the Hessian $\tens{\tens{H}}$ that implicitly enter Eqs.~\eqref{9ampl} (see Appendix~\ref{app-a} for details) are
\begin{equation}\label{eq:H_cmp}
\frac{\partial L_{22}(k^2)}{\partial \beta} = -k^2\,, \qquad
H_{111}(k^2) = k^2  f_{uu} = -k^2  6u_s\,, \qquad
H_{222}(k^2) = k^2  g_{vv} = -k^2  6v_s\,.
\end{equation}
Note that for the special case of a trivial homogeneous state, i.e., $u_s=v_s=0$, quadratic interactions are absent and the description of resonances via the leading order amplitude equations~\eqref{9ampl} does not apply.

Incorporating the imaginary part of $\mu_o$ into the frequency, the coefficients in Eqs.~\eqref{9ampl} become
\begin{align}
\begin{split}\label{eq:coeff_final}
\mu_+ =&\varepsilon  \frac{\left(L^2 \left(1-3 u_s^2\right)-16 \pi ^2\right)}{12 \pi ^2 (\sigma +1)}\,, \quad \mu_o = -\frac{\varepsilon}{2}\,, \quad \nu_+ = \frac{L^2 (v_s-u_s (3 u_s v_s+1))-16 \pi ^2 v_s}{2 \pi ^2 (\sigma +1)}\,, \cr
 \nu_o =&\frac{3 L^2 u_s}{L^2 \left(3 u_s^2-1\right)+16 \pi ^2}-3 v_s \cr
 +& \I \frac{12 \pi ^2 \left(L^2 \left(3 u_s^2-1\right)+4 \pi ^2\right) \left(L^2 \left(3 u_s^2 v_s+u_s-v_s\right)+16 \pi ^2 v_s\right)}{\left(L^2 \left(3 u_s^2-1\right)+16 \pi ^2\right) L^4 \omega _o}\,.
 \end{split}
\end{align}
\begin{figure}[t]
\centering
\includegraphics[width=0.4\hsize]{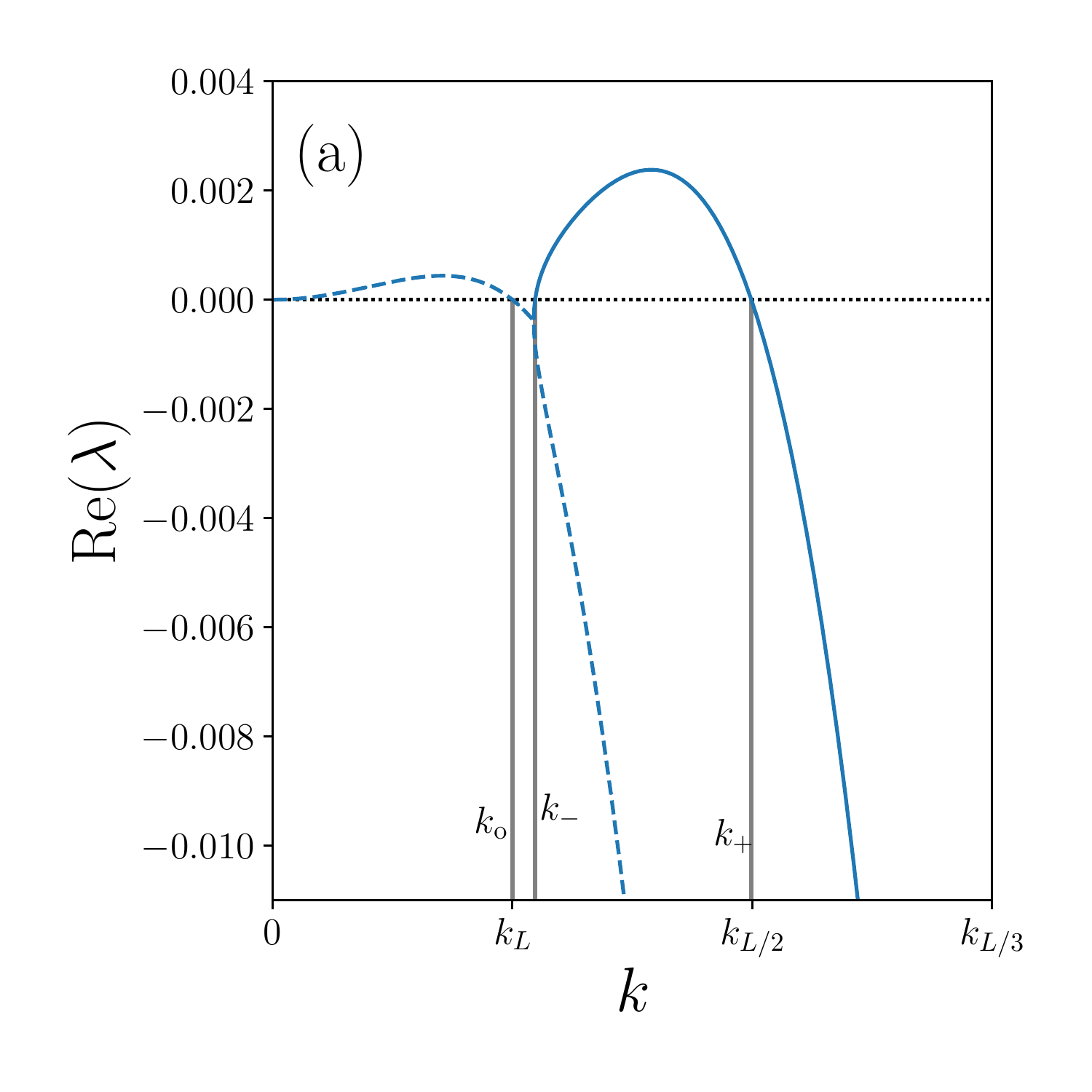}
\includegraphics[width=0.55\hsize]{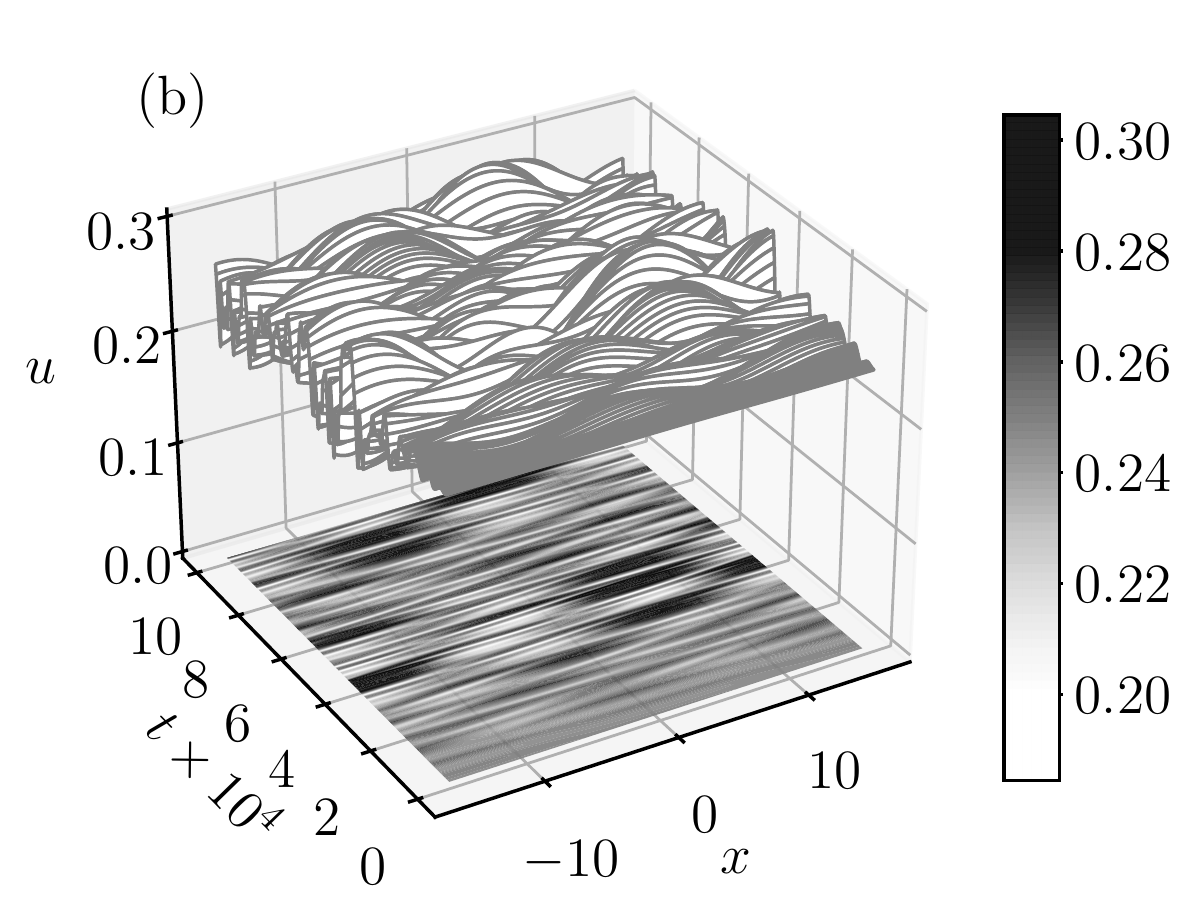}
\caption{(a) Dispersion relation $\mathrm{Re}\,\lambda(k)$ [Eq.~\eqref{eq:lambda}] for a finite system of the size $L$ at parameter values where there are resonant marginal modes $k_\text{o}=k_L$ [Eq.~\eqref{eq:ko}] and $k_+=k_{L/2}$ [Eq.~\eqref{eq:kpm}] of the conserved-Hopf and conserved-Turing instabilities, respectively. (b) Space-time plot of $u(x,t)$ obtained by fully nonlinear time simulation of  the mass-conserving modified FitzHugh--Nagumo system showing two-frequency oscillatory behavior. The parameters are $\sigma \approx 1.19557>1$, $f_u[=1-3u_s^2]\approx 0.81867$, $f_v=-1$, $g_u\left[=\alpha\right]\approx  0.60739$ and $g_v\left[=-\beta-3v_s^2\right]\approx -0.73075$ where for the specific model $\beta = \beta_c + \varepsilon \approx 0.15016$ with $\varepsilon=-10^{-4}$, $u_s \approx 0.24585$ and $v_s \approx 0.43992$. The domain size is $L=10 \pi$. In the formulation of the amplitude equations~\eqref{1amplreal}-\eqref{1amplim}, the corresponding parameters are $\mu= 0.05$ and $\varphi = -0.65 \pi$.}\label{fig:oscillation}
\end{figure}

We have performed direct time simulations of this model in the vicinity of the degenerate bifurcation where we expect resonance behavior. In particular, we choose $\mu_o>0$ and $\mu_+<0$ as in section~\ref{sec:resonance}, i.e., we decrease $\beta$ by $\varepsilon<0$. 
Fixing a specific domain size $L$, there are $\sigma,\, u_s$ and $v_s$ as free parameters of the original model that can be used to adjust three parameters of Eqs.~\eqref{eq:coeff_final}. To compare the specific model to  the weakly nonlinear results in Fig.~\ref{fig:f1wt} [from Eqs.~\eqref{1amplreal} and \eqref{1amplim}] we adjust $\mu = \mu_o k_o^2/(|\mu_+| k_+^2)$, and $\varphi$ as the phase of the complex parameter $\nu_o$. As there is an additional free parameter in the original model we have further fixed the absolute value of $\nu_o$ to one. Note that in the absence of a cubic nonlinearity in $g(u,v)$, i.e., for the fully mass-conserving version of the standard Fitz--Hugh Nagumo model, $H_{222}(k^2)$ in Eq.~\eqref{eq:H_cmp} vanishes and $v_s$ does not contribute to the coefficients in Eqs.~\eqref{eq:coeff_final}. In principle, there are still enough free parameters to adjust $\mu$ and $\varphi$, practically however, we are unable to enter the parameter range where we expect oscillatory states, i.e., the range depicted in Fig.~\ref{fig:f1wt}~(b). This is due to further restrictions based on the various inequalities involved in the occurrence of the instabilities (cf.~Eqs.~\eqref{eq:restrictions}) and the considered signs in the amplitude equations~\eqref{1amplreal} and \eqref{1amplim} (cf.~Appendix~\ref{app-a}). Including the cubic nonlinearity in $g(u,v)$ solves this issue.
A typical result is given in
Fig.~\ref{fig:oscillation} where panel~(a) gives the dispersion relation at parameter values where $k_\text{o}=k_L$ and $k_+=k_{L/2}$ are resonant, while panel~(b) shows a space-time plot of the corresponding time simulation. The latter indeed shows a two-frequency behavior analogous to the secondary oscillations found with the weakly nonlinear approach in section \ref{sec:resonance} (cf.~Fig.~\ref{fig:f1wt}).

However, performing time simulations at parameters that correspond to regions ``A'' and ``S'' in Fig.~\ref{fig:f1wt}~(b) where asymmetric and symmetric stationary solutions of the amplitude equations \eqref{9ampl} are respectively stable, we encounter  only traveling waves (TW), standing waves (SW) (both with wavenumber $k_L$), or stationary Turing patterns (ST) with the wavenumber $k_{L/2}$. Due to the scaling in the employed ansatz, these states are not captured by the amplitude equations.

\section{Summary and outlook}
\label{sec:conc}

We have considered a nonreciprocally coupled two-field Cahn-Hilliard system that is known to allow for oscillatory behavior and  suppression of coarsening.  We have reviewed the linear stability analysis of uniform steady states and have shown that for general intra- and interspecies interaction terms all instability thresholds of the fully mass-conserving Cahn-Hilliard system are identical to the ones for the corresponding nonmassconserving reaction-diffusion system. Next, we have briefly highlighted the differences in the linear behavior of conserved and nonconserved models that occur beyond the instability onset. Focusing on the codimension-two point where conserved-Hopf and conserved-Turing instabilities simultaneously occur, we have discussed possible interactions of linear modes. In particular, we have analyzed the specific case of a ``Hopf-Turing'' resonance. To do so, we have first employed a weakly nonlinear approach to consider the amplitude equations close to the codimension-two point. After discussing the  behavior of solutions in the general case, we have derived the coefficients of the amplitude equations for a specific nonreciprocal Cahn-Hilliard model that corresponds to a modified conserved  FitzHugh-Nagumo model. Although a conserved version of the standard FitzHugh-Nagumo model shows a codimension-two point, it does not allow to adjust parameters in such a way that the parameter ranges of the weakly nonlinear model corresponds to the interesting cases shown in Fig.~\ref{fig:f1wt}~(b). This may be due to the fact that it is a nongeneric model, see discussion in Ref.~\cite{FHKG2022arxiv}. Finally, we have shown that fully nonlinear time simulations indeed show two-frequency behavior analogous to the secondary oscillations discussed in the framework of the weakly nonlinear theory.

However, we have also noted that not all behavior predicted by the amplitude equation is found in the fully nonlinear calculation. To better understand where weakly and fully nonlinear results agree and where they disagree the mapping of the respective parameter sets and resulting behavior should be further scrutinized in the future. A problem that needs further attention is that the parameter mapping is not one-to-one and itself is highly nonlinear. This makes it, for instance, quite difficult to identify parameter ranges where certain states dominate in a nonlinear model with corresponding ranges in the weakly nonlinear description. The usage of continuation methods might allow one to obtain bifurcation diagrams for nonlinear models that could be then directly compared to a bifurcation diagram presented in the weakly nonlinear case. This would also allow one to clarify the question whether an additional inclusion of cubic terms into the weakly nonlinear approach has a major impact.

\acknowledgments
TFH and UT acknowledge support from the doctoral school ``Active Living Fluids'' funded by the German-French University (Grant No. CDFA-01-14). TFH was supported by the foundation ``Studienstiftung des deutschen Volkes''. The authors thank Svetlana V. Gurevich for fruitful discussions.

\appendix
\section{Details of weakly nonlinear analysis}\label{app}
%
 \subsection{Derivation of the amplitude equations~\eqref{1amplreal}-\eqref{1amplim}}\label{app-a}
In this section we provide more details on the calculations discussed in section~\ref{sec:resonance} of the main text. We consider the system \eqref{91uwas} close to a wave-Turing resonance, where the three wavevectors form an isosceles triangle. The steady state is denoted by  $\vec u_s$. In one spatial dimension (1D), the ``triangle'' is flat and the wavenumber of the oscillatory mode is simply half the wavenumber of the stationary one. To be definite, we concentrate here on this case, though all derivations apply to a general situation. In order to capture nonlinear interactions by a weakly nonlinear approach, we further demand that both corresponding growth rates are small, i.e., we are close to the codimension-two bifurcation that occurs in an infinite system under the conditions given by Eq.~\eqref{eq:a22cd2}. In a specific finite-size system of length $L$ the condition is modified to $2 k_o= k_+=k_{L/2}=4\pi/L$  where $k_o$ [Eq.~\eqref{eq:ko}] and $k_+$ [Eq.~\eqref{eq:kpm}] correspond to the marginal modes of the conserved-Hopf and the conserved-Turing instability, respectively. Note that the following results equivalently apply if one chooses the stationary marginal mode with $k_-$ instead of $k_+$ to be in resonance with the oscillatory ones. The two conserved-Hopf modes correspond to left and right traveling waves in 1D. In an isotropic system they exhibit the same dispersion, i.e., they have the same frequency and the same eigenvector.
Two parameters have to be set to specific critical values that adjust the codimension-two point. 
Then, we use one of them, say $\beta$, and introduce a small deviation, i.e., $\beta=\beta_\mathrm{c}+ \varepsilon$.
Since $\varepsilon\ll 1$, it can be used as a small parameter, and we expand all fields in $\varepsilon$ as
\begin{align}
\begin{split}\label{eq:app:ansatz}
\vec u =& \vec u_s + \varepsilon \vec u_1(x,t,T)+ \varepsilon^2 \vec u_2(x,t,T) +\mathcal{O}(\varepsilon^3) ~\\
\text{with}\,\,\, \vec u_1(x,t,T) =&a_+(T) \vec u_+ e^{\text{i}  k_+   x} +a_{o1}(T) \vec u_{o} e^{\text{i} \left(-k_{o}  x + \omega_{o} t\right)} +a_{o2}(T) \vec u_{o} e^{\text{i} \left(k_{o} x + \omega_{o} t\right)} + \text{c.c.} \,.
\end{split}
\end{align}
The amplitudes $a_+(T)$, $a_{o1}(T)$, $a_{o2}(T)$  evolve on a large timescale $T=\varepsilon t$  and correspond to the stationary, right-traveling and left-traveling mode, respectively, with their corresponding zero eigenvectors $\vec u_+ \in \mathbb{R}$ and $\vec u_{o}\in \mathbb{C}$. 
The frequency $\omega_{o}$ is the imaginary part of the eigenvalues at onset of instability of the wave modes. Note that Eq.~\eqref{eq:app:ansatz} results from the more general ansatz~\eqref{eq:ansatz} in the main text if isotropy is used. 
We introduce \eqref{eq:app:ansatz} into \eqref{91uwas} and expand in $\varepsilon$ to obtain
\begin{equation}\label{eq:app:2}
\varepsilon^2 \left(\partial_T \vec u_1(x,t,T) + \tens{\mathcal{ L}} \cdot \vec u_2(x,t,T) \right) + \mathcal{O}(\varepsilon^3) = \varepsilon^2 \left( \frac{\partial \tens{L}}{\partial \beta}(\beta-\beta_c) \cdot \vec u_1 + \vec{u}_1 \cdot \tens{\tens{H}} \cdot \vec u_1\right)+\mathcal{O}(\varepsilon^3), 
\end{equation}
where 
\begin{equation}
\tens{\mathcal{ L}}= \tens{\mathbf{1}} \partial_t - \tens{L}=
\tens{\mathbf{1}} \partial_t + \partial_{xx}\left(
\begin{array}{c c}
\partial_{xx} +f_u  &  f_v ~\\
  g_u &  \partial_{xx} +g_v ~\\
\end{array}\right)
\end{equation} 
is the linear partial differential operator that includes both the time derivative with respect to the fast timescale $t$ and the Jacobian matrix $\tens{L}$, written here in spatial representation.
$\tens{\mathbf{1}}$ is the unit matrix and the zero eigenmodes solve
\begin{equation}
\tens{\mathcal{L}} \cdot \vec u_+ e^{\I k_+ x}=\tens{\mathcal{L}} \cdot \vec u_o e^{\I(\pm k_o x + \omega_o)}=\vecg 0\,.
\end{equation}
Further, we define the Hessian  
$\tens{\tens{H}} = \nabla_{\vec u} \tens{L}$.
Both $\tens{\mathcal{ L}}$  and  $\tens{\tens{H}}$ are evaluated for the uniform steady state $\vec u= \vec u_s$ at $\beta=\beta_c$.
The special property of the Hopf-Turing resonance is that quadratic terms are sufficient to obtain a saturated system, so that, for our purpose, we neglect all higher terms.
Next, we multiply the remaining $\mathcal{O}(\varepsilon^2)$ terms in Eq.~\eqref{eq:app:2} by one of the three adjoint linear modes that solve the adjoint linear eigenvalue problem, i.e.,
\begin{equation}
\vec u_+^{\dagger} e^{-\I k_+ x}\cdot \tens{\mathcal{L}}  =\vec u_o^\dagger e^{-\I(\pm k_o x + \omega_o)} \cdot \tens{\mathcal{L}}  =\vecg 0\,,
\end{equation}
and integrate over the whole domain.
In each case, the term that involves $\vec u_2$ in Eq.~\eqref{eq:app:2} vanishes and the integration amounts to a projection on the corresponding Fourier modes. We normalize all adjoint vectors via $\vec u_+^{\dagger} \cdot \vec u_+ =\vec u_o^\dagger\cdot \vec u_o=1$. This gives
\begin{align}
\begin{split}\label{eq:app:AE}
e^{\I k_+ x} \sim \,\, \,\,\partial_T a_+ = &\vec u_+^\dagger \cdot \frac{\partial \tens{L}(k_+^2)}{\partial \beta}\cdot\vec{u}_+ (\beta-\beta_c) a_+ +  \vec u_+^\dagger \cdot \left(\vec{\bar u}_o \cdot \tens{\tens{H}}(k_+^2) \cdot \vec u_o\right) \bar a_{o1}  a_{o2}\,,~\\
e^{\I(-k_o x+ \omega_o t)} \sim \,\, \, \partial_T a_{o1} = &\vec u_o^\dagger \cdot \frac{\partial \tens{L}(k_o^2)}{\partial \beta}\cdot\vec{u}_o (\beta-\beta_c) a_{o1} +  \vec u_o^\dagger \cdot \left(\vec{u}_+ \cdot \tens{\tens{H}}(k_o^2) \cdot \vec u_o\right) \bar a_{+}  a_{o2}\,,~\\
e^{\I(k_o x+ \omega_o t)} \sim \,\, \,\, \partial_T a_{o2} = &\vec u_o^\dagger \cdot \frac{\partial \tens{L}(k_o^2)}{\partial \beta}\cdot\vec{u}_o (\beta-\beta_c) a_{o2} +  \vec u_o^\dagger \cdot \left(\vec{u}_+ \cdot\tens{\tens{H}}(k_o^2) \cdot \vec u_o\right) a_{+}  a_{o1}\,,~
\end{split}
\end{align}
where we use the resonance condition
\begin{equation}
e^{\I k_+ x}e^{\I(-k_o x+ \omega_o t)} = e^{\I((k_+-k_o) x+ \omega_o t)}=e^{\I(k_o x+ \omega_o t)}\,,
\end{equation}
that gives the quadratic coupling terms. All quantities with a bar denote complex conjugates.
In the resulting system of amplitude equations \eqref{eq:app:AE} all spatial derivatives are replaced by the corresponding wavenumber. 
Then Eqs.~\eqref{eq:app:AE} is rewritten as
 \begin{align} 
 \begin{split}\label{app:9ampl}
\dot a_+ &= k_+^2 (\mu_+ a_+ + \nu_+ \bar a_{o1} a_{o2} )\,,~\\
\dot a_{o1} & = k_o^2(\mu_o a_{o1} + \nu_o \bar a_{+} a_{o2} )\,, \quad 
\dot a_{o2}  =  k_o^2(\mu_o a_{o2} + \nu_o a_+  a_{o1})\,,
\end{split}
\end{align}
where the coefficients $\mu_+, \, \nu_+ \in \mathbb{R}$ and $\mu_o, \, \nu_o \in \mathbb{C}$ resemble the structure of the amplitude system in the nonconserved case \cite{PiRu1999csf}, but in the conserved case additional prefactors consisting of squared wavenumbers occur. We apply the transformation $ \tilde a_{o1/2} =  a_{o1/2} e^{-\I k_o^2 \Im \mu_o \, t}$, i.e.,
\begin{equation}
\dot{\tilde a}_{o1/2} = \left(\dot a_{o1/2} -\I k_o^2 \Im \mu_o \, a_{o1/2} \right) e^{-\I k_o^2 \Im \mu_o \, t}\,,
\end{equation}
that eliminates the contribution of the imaginary part of $\mu_o$ in Eqs.~\eqref{app:9ampl}. 
In the following, we omit the tilde, and consider $\mu_o$ as real. Next, we introduce a polar representation of the complex amplitudes, $a_+ = \rho_+\; \E^{\I \theta_+}, \; a_{o1} = \rho_1 \E^{\I \theta_1}, \; a_{o2} = \rho_2 \E^{\I \theta_2}$ and of the remaining complex coefficient $\nu_o = \nu \E^{\I \varphi}$ where, by construction, $\rho_+, \rho_1, \rho_2, \nu >0$. Then Eqs.~\eqref{app:9ampl} become
 \begin{align} 
 \begin{split}\label{app:10ampl}
\left(\dot \rho_+ + \I \dot \theta_+ \right) \E^{\I \theta_+} &= k_+^2 \left(\mu_+ \rho_+ \E^{\I \theta_+} + \nu_+ \rho_{1} \rho_{2} \E^{\I (\theta_2 - \theta_1)} \right)\, ,\\
\left(\dot \rho_1 + \I \dot \theta_1 \right) \E^{\I \theta_1} & = k_o^2\left(\mu_o \rho_{1}\E^{\I \theta_1}+ \nu \rho_+  \rho_{2} \E^{\I (-\theta_+ + \theta_2 + \varphi)}\right)\,,~\\
\left(\dot \rho_2 + \I \dot \theta_2 \right) \E^{\I \theta_2}  & =  k_o^2\left(\mu_o \rho_{2}\E^{\I \theta_2} + \nu \rho_+  \rho_{1} \E^{\I (\theta_+ + \theta_1 + \varphi)}\right)\,.
\end{split}
\end{align}
We divide by the respective exponential factor on each left-hand side and introduce the relative phase $\Theta = \theta_+ + \theta_1 - \theta_2$ that can be identified as the only relevant phase information that enters the dynamics. The real and imaginary parts of Eqs.~\eqref{app:10ampl} give the dynamics of the corresponding real amplitude and phase, respectively.
Further, the dynamics of the phases are combined to give $\dot \Theta$. Then the amplitude equations of the relevant field quantities are
\begin{align} 
\dot{\rho}_+ & =  k_+^2 \left(\mu_+ \rho_+ + \nu_+ \rho_1 \rho_2 \cos \Theta\right)\,, \cr
\dot{\rho}_1 &= k_o^2 \left( \mu_o \rho_1 +  
     \nu \rho_+ \rho_2 \,\cos (\Theta - \varphi)\right)\,,\quad
 \dot{\rho}_2 = k_o^2 \left( \mu_o \rho_2 +  
     \nu \rho_+ \rho_1 \,\cos (\Theta + \varphi)\right)\,,~\label{app:11}\\
\dot{\Theta} &=  \dot \theta_+ + \dot \theta_1 - \dot \theta_2 =  - k_+^2 \nu_+ \frac{\rho_1 \rho_2}{\rho_+} \sin \Theta - \nu k_o^2 \rho_+ \left(\frac{\rho_1}{\rho_2}\sin(\Theta+\varphi) + \frac{\rho_2}{\rho_1}\sin(\Theta-\varphi)\right)\,.
\notag\end{align} 
 It follows from the first equation in \eqref{app:11} that it is necessary for the existence of stationary solutions that either $\mu_+$ and $\nu_+$ have the same sign and $\cos \Theta<0$, i.e., $|\Theta|>\pi/2$, or they have opposite sign, implying $\cos \Theta>0$, i.e., $|\Theta|<\pi/2$. Here we assume $\nu_+>0, \, \mu_+<0$, the latter implying that the Turing mode is linearly weakly damped. We apply a rescaling to eliminate all but two effective parameters. Specifically, we use $1/(k_+^2 |\mu_+|)$ as the time scale, $|\mu_+|k_+^2/(\nu k_o^2)$ as the scale of $\rho_+$, and $|\mu_+|k_+/(\sqrt{\nu \nu_+}k_o)$ as the scale of $\rho_{1/2}$. It reduces Eqs.~\eqref{app:11} to
\begin{align} \begin{split} \label{app:12} 
\dot{\rho}_+ & =  -\rho_+ + \rho_1 \rho_2 \cos \Theta\,, \quad
\dot{\rho}_1 = \mu \rho_1 +  
     \rho_+ \rho_2 \,\cos (\Theta - \varphi)\,, \quad
\dot{\rho}_2 = \mu \rho_2 +  
     \rho_+ \rho_1 \,\cos (\Theta + \varphi)  \,, \cr
\dot{\Theta} & =  - \frac{\rho_1 \rho_2}{\rho_+} \sin \Theta
  -  \rho_+ \left(\frac{\rho_1}{\rho_2}\sin(\Theta+\varphi)+ \frac{\rho_2}{\rho_1}\sin(\Theta-\varphi) \right)\,,
     \end{split} \end{align} 
where $\mu = \mu_o k_o^2/(|\mu_+| k_+^2)$ and $\varphi$ are the two remaining free parameters.
\subsection{Stationary solutions}\label{app-b}
Setting the time derivatives to zero defines stationary solutions. The first three equations of \eqref{app:12} define the stationary values of the amplitudes
\begin{equation}
{\rho}_+ = \frac{|\mu|}{[\cos(\Theta-\varphi)\cos(\Theta+\varphi)]^{1/2}}\,, \quad
{\rho}_{1,2} = \left[-\frac{\mu}{\cos(\Theta \pm\varphi)\cos \Theta}\right]^{1/2}\,. 
\label{app:13}    \end{equation}
All amplitudes need to be positive and, hence, it follows from \eqref{app:13} that $\cos(\Theta \pm \varphi)<0$, i.e., $\pi/2<|\varphi|<\pi$. 
Using (\ref{app:13}) in the equation for $\Theta$ in (\ref{app:12}) defines the stationary values of $\Theta$ by
\begin{equation} 
- \tan \Theta+\mu[ \tan (\Theta-\varphi)+ \tan (\Theta+\varphi)]= 0\,.
      \label{app:14}  \end{equation} 
The trivial solution of \eqref{app:14} is $\Theta =0$ which is the symmetric stationary solution, since it follows from Eqs.~\eqref{app:13} that $\rho_+ = \rho_1^2 = \rho_2^2 = |\mu/\cos \varphi |$, i.e., left and right traveling wave modes have the same amplitude.
Nonzero $\Theta$ corresponds to asymmetric solutions. We use the identity
\begin{align}
\begin{split}
\tan (\Theta-\varphi)+ \tan (\Theta+\varphi) = & \frac{\sin \Theta \cos \varphi - \cos \Theta \sin \varphi}{\cos \Theta \cos \varphi + \sin \Theta \sin \varphi} +  \frac{\cos \Theta \sin \varphi + \sin \Theta \cos \varphi}{\cos \Theta \cos \varphi - \sin \Theta \sin \varphi} ~\\
=& \frac{2 \cos \Theta \sin \Theta }{\cos^2 \Theta \cos^2 \varphi - \sin^2 \Theta \sin^2 \varphi}
\end{split}
\end{align} 
and transform \eqref{app:14} into
\begin{align}
\begin{split}\label{app:angle}
- \tan \Theta+\mu \frac{2 \cos \Theta \sin \Theta }{\cos^2 \Theta \cos^2 \varphi - \sin^2 \Theta \sin^2 \varphi} = & 0 ~\\
\Rightarrow (1- \cos^2 \Theta) \sin^2 \varphi -\cos^2 \Theta \cos^2 \varphi + 2 \mu \cos^2 \Theta = & 0~\\
\Rightarrow (1-2 \mu) \cos^2 \Theta = & \sin^2 \varphi\,.
 \end{split}
\end{align}
Since $(1-2 \mu) \cos^2 \Theta< 1-2 \mu$, asymmetric solutions can only exist if $1-2 \mu\geq \sin^2 \varphi $, i.e., for $0\leq \mu \leq \frac12 \cos^2\varphi$. The acceptable interval of both angles, as well as the sign of $\mu$ would overturn if we had chosen $ \mu_+>0$.
\subsection{Stability of stationary solutions}\label{app-c}
Next, we consider the stability of the symmetric and asymmetric solutions.
For this we determine the Jacobian matrix $\tens J$,  
\begin{align}
\begin{split}
\tens J = \left(\begin{array}{c c c c}
-1 & \rho_2 \cos \Theta & \rho_1 \cos \Theta & - \rho_1 \rho_2 \sin \Theta~\\
\rho_2 \cos(\Theta - \varphi) & \mu & \rho_+ \cos(\Theta - \varphi)& -\rho_+ \rho_2 \sin(\Theta - \varphi)~\\
\rho_1 \cos(\Theta + \varphi)&  \rho_+ \cos(\Theta - \varphi) & \mu & -\rho_+ \rho_1 \sin(\Theta - \varphi)~\\
\frac{P - Q  -R }{\rho_+^2 \rho_1 \rho_2} 
&
\frac{- P +Q - R }{\rho_+ \rho_1^2 \rho_2} 
&
\frac{-P - Q +R}{\rho_+ \rho_1 \rho_2^2} 
&
-\frac{S}{\rho_+ \rho_1 \rho_2} 
\end{array}\right)
\end{split}
\end{align}
with abbreviations 
\begin{align}
P&= \rho_1^2 \rho_2^2 \sin \Theta, \quad Q = \rho_+^2 \rho_2^2\sin(\Theta- \varphi), \quad 
R=\rho_+^2 \rho_1^2\sin(\Theta+ \varphi), \cr
S&=\rho_1^2 \rho_2^2 \cos \Theta + \rho_+^2 \rho_2^2 \cos(\Theta- \varphi) +\rho_+^2 \rho_1^2 \cos(\Theta+ \varphi)
\end{align}
Plugging the stationary solution, we compute the characteristic polynomial
\begin{equation}
\det \left( \lambda \tens{\mathbf 1}-\tens J  \right)=  \lambda^4 + a_1 \lambda^3 + a_2 \lambda^2 + a_3 \lambda + a_4=0
\end{equation} with constant real coefficients $a_i$.
We apply the Hurwitz criterion to analyze linear stability. For linear stability the following conditions have to be fulfilled
\begin{equation}\label{app:criterion}
a_1>0, \quad a_4>0, \quad a_1 a_2 - a_0 a_3 >0, \quad
(a_1 a_2 - a_0 a_3) a_3 - a_1^2  a_4 >  0\,.
\end{equation}
In particular, zero crossings of $a_4$ that is the determinant of  $\tens J$ indicate monotonic instabilities.

\begin{figure}
\includegraphics[width=0.33\textwidth]{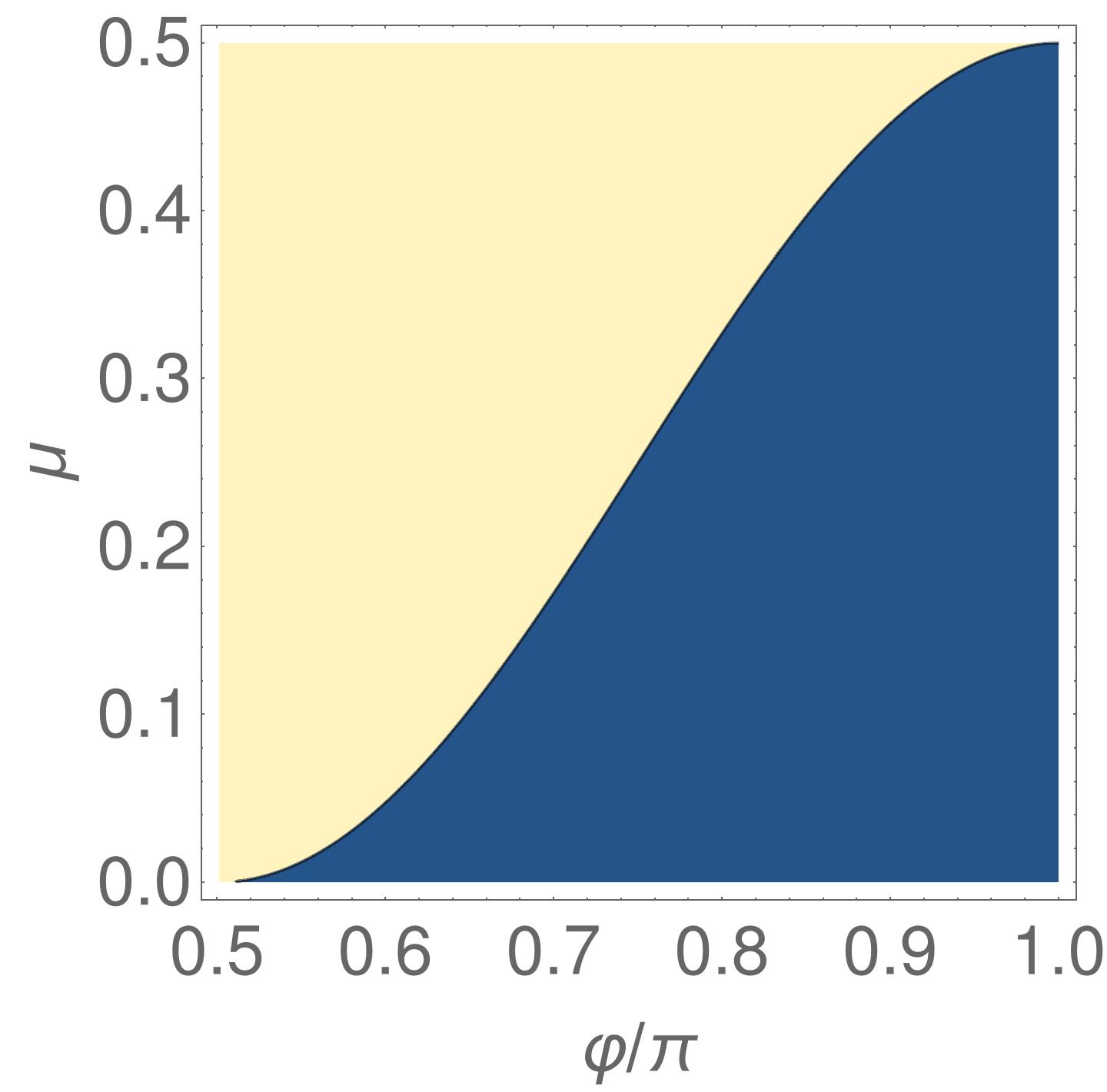}~\hfill
\includegraphics[width=0.33\textwidth]{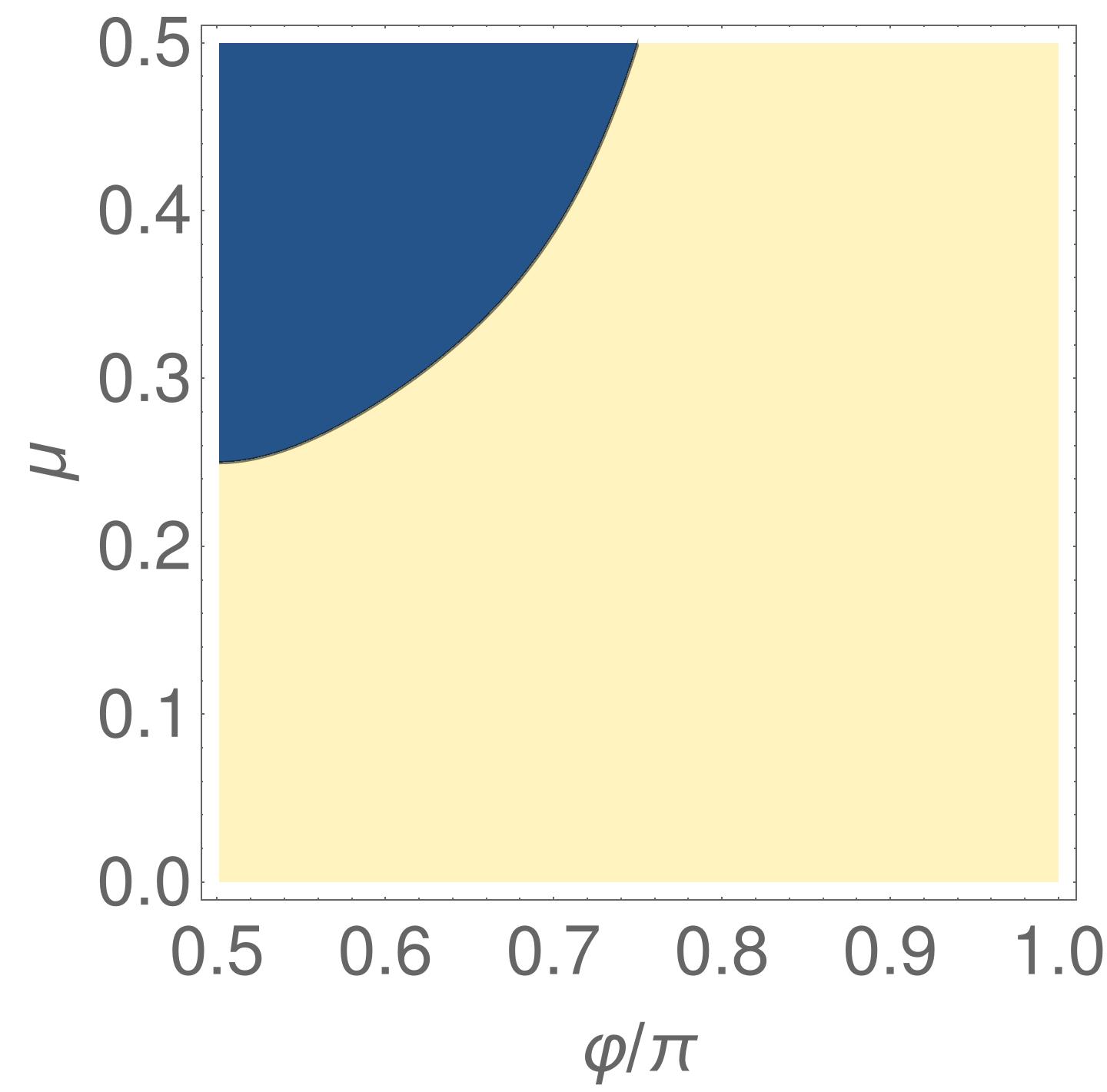}~\hfill
\includegraphics[width=0.33\textwidth]{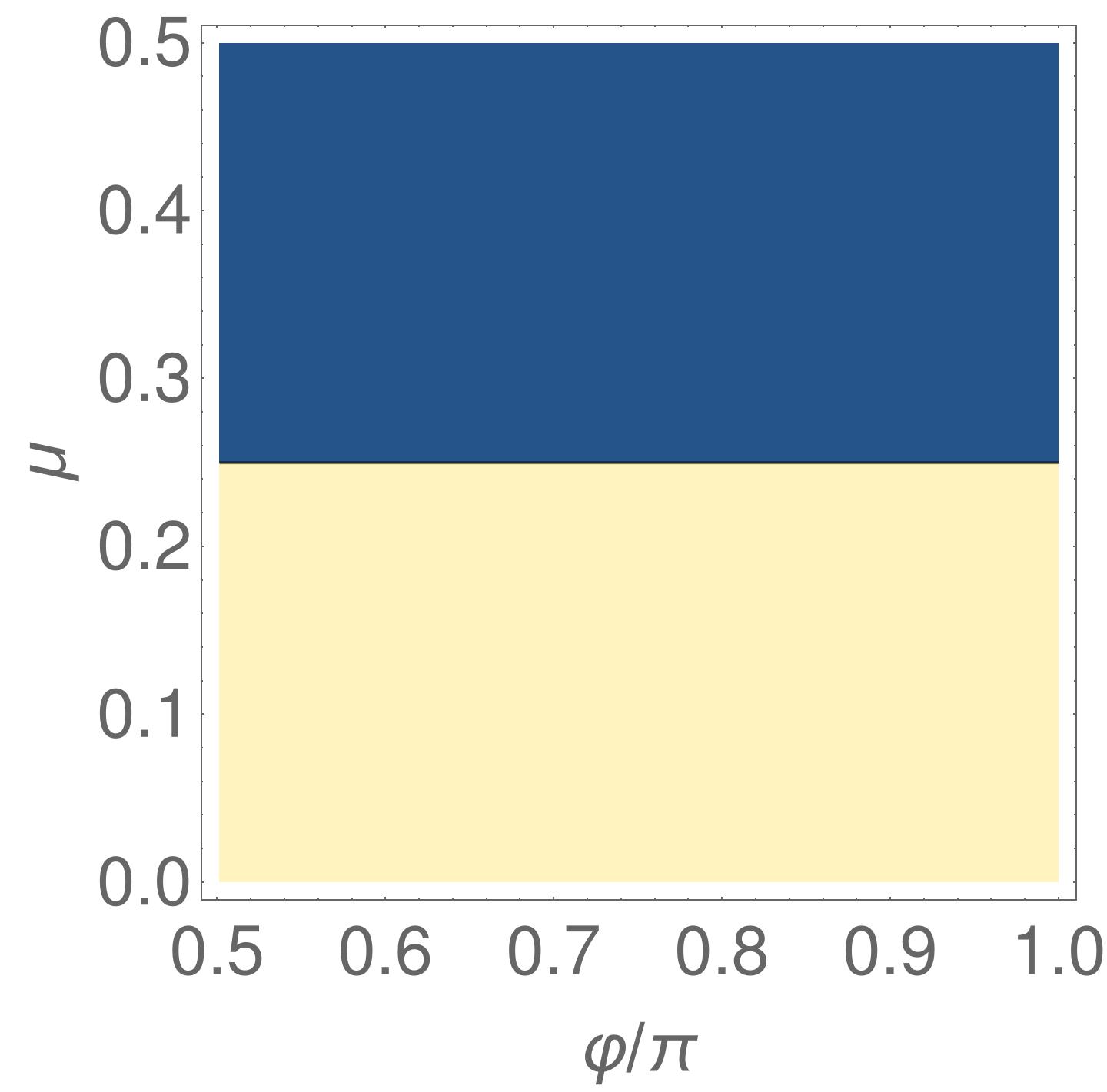}~
\caption{Hurwitz criterion for the symmetric solution, Eqs.~(a)~\eqref{app:Hsym_2}, (b)~\eqref{app:Hsym_3} and (c)~\eqref{app:Hsym_4}. Yellow [blue] regions correspond to positive [negative] values. }\label{app:fig:Hcrit_sym}
\end{figure}

For the symmetric solution the Jacobian matrix is
\begin{equation}
\tens J_{\text{sym}}= \left(\begin{array}{c c c c}
-1 & \sqrt{\frac{-\mu}{\cos \varphi}} & \sqrt{\frac{-\mu}{\cos \varphi}}  & 0~\\
\sqrt{\frac{-\mu}{\cos \varphi}}  & \mu & -\mu & \left(\frac{-\mu}{\cos \varphi} \right)^{\frac32} \sin \varphi ~\\
\sqrt{\frac{-\mu}{\cos \varphi}}  & -\mu & \mu &  -\left(\frac{-\mu}{\cos \varphi} \right)^{\frac32} \sin \varphi~\\
0 & -2 \sqrt{\frac{-\mu}{\cos \varphi} }\sin \varphi & 2 \sqrt{\frac{-\mu}{\cos \varphi} }\sin \varphi  & -1 +2\mu
\end{array}\right)
\end{equation}
and the
Hurwitz criterion yields
\begin{align}
2- 4\mu>&0\,,~\label{app:Hsym_1} \\
4 \mu^2 (-1 +2 \mu + 2 \mu \tan^2\varphi)>&0\,,~ \label{app:Hsym_2} \\
2 - 12 \mu + 28 \mu^2 - 16 \mu^3 - 4 \mu^2 (-1 + 4 \mu) \tan^2 \varphi >&0\,,~\label{app:Hsym_3} \\
-4 \mu^2 (-1 + 4 \mu) (1 - 3 \mu + 4 \mu^2 + (1 - 2 \mu) \cos(2 \varphi) + 
   \mu \cos(4 \varphi)) \sec^4 \varphi > & 0\,.\label{app:Hsym_4}
\end{align}
In the following, we always consider the region $0\leq \mu \leq 1/2$, $\pi/2\leq \varphi \leq \pi$. Note that all results are also valid for $\varphi \to - \varphi$. 
Eq.~\eqref{app:Hsym_1} holds for the considered parameter region.
The other three conditions are plotted in Fig.~\ref{app:fig:Hcrit_sym} where the yellow regions in panels~(a), (b)  and (c) correspond to positive values of the expression in Eqs.~\eqref{app:Hsym_2}, \eqref{app:Hsym_3} and \eqref{app:Hsym_4}, respectively. 
Together they yield $\left[\frac12 \cos^2 \varphi, \frac14 \right]$ as the region of linear stability for the symmetric solution. At $\mu = \frac12 \cos^2 \varphi$ where $a_4$ crosses zero, the asymmetric solution emerges from the symmetric one in a pitchfork bifurcation and the latter is unstable for smaller $\mu$ values. At $\mu=\frac14$ a Hopf bifurcation renders the symmetric solution unstable for larger $\mu$ values. At $\varphi=\frac34 \pi$, i.e., when $\frac12 \cos^2 (\frac34 \pi)= \frac14$ the pitchfork and the Hopf bifurcation merge at the double zero singularity and no stable symmetric solutions exist for $\varphi\geq \frac34 \pi$.
\begin{figure}
\includegraphics[width=0.33\textwidth]{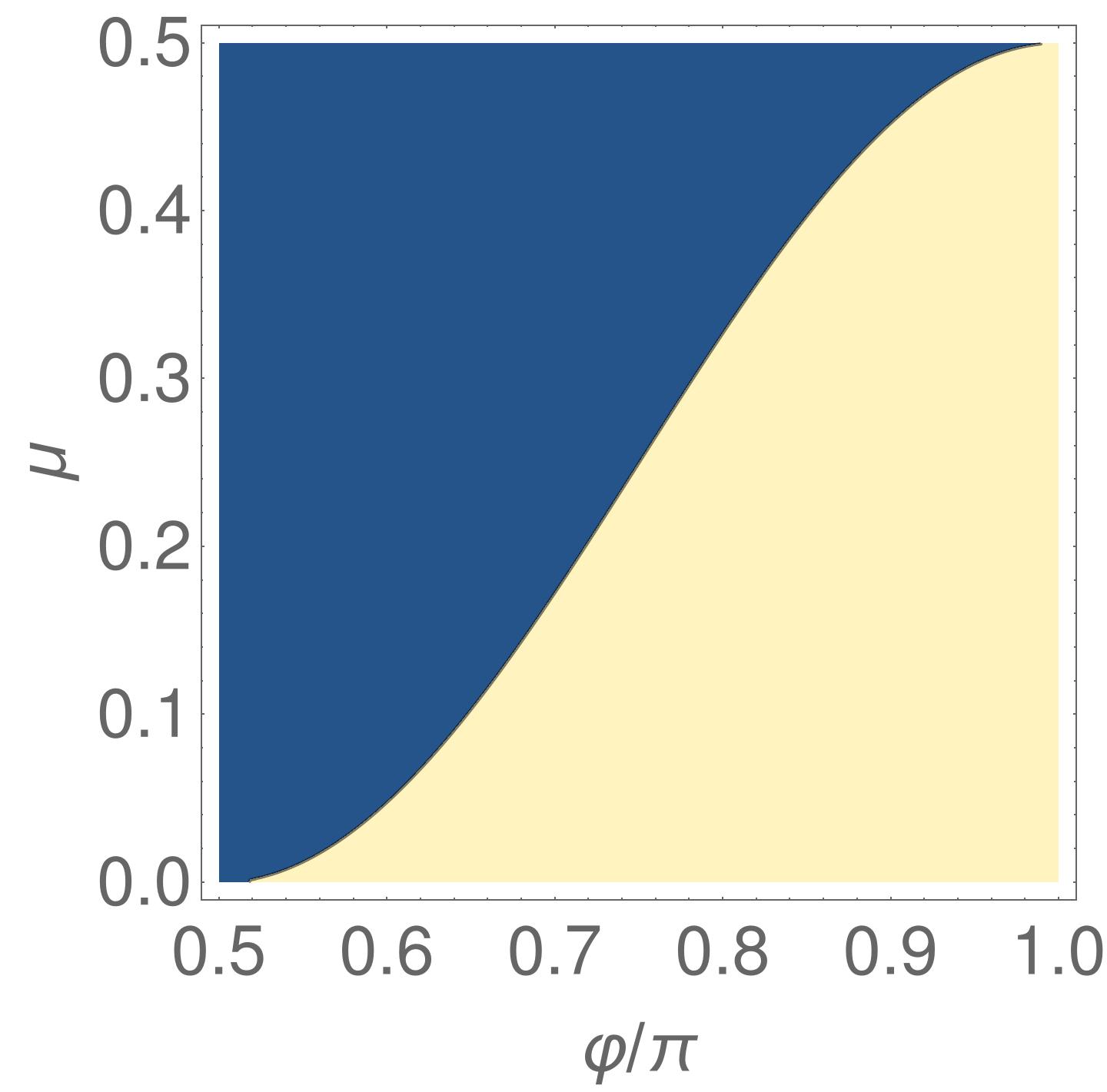}~\hfill
\includegraphics[width=0.33\textwidth]{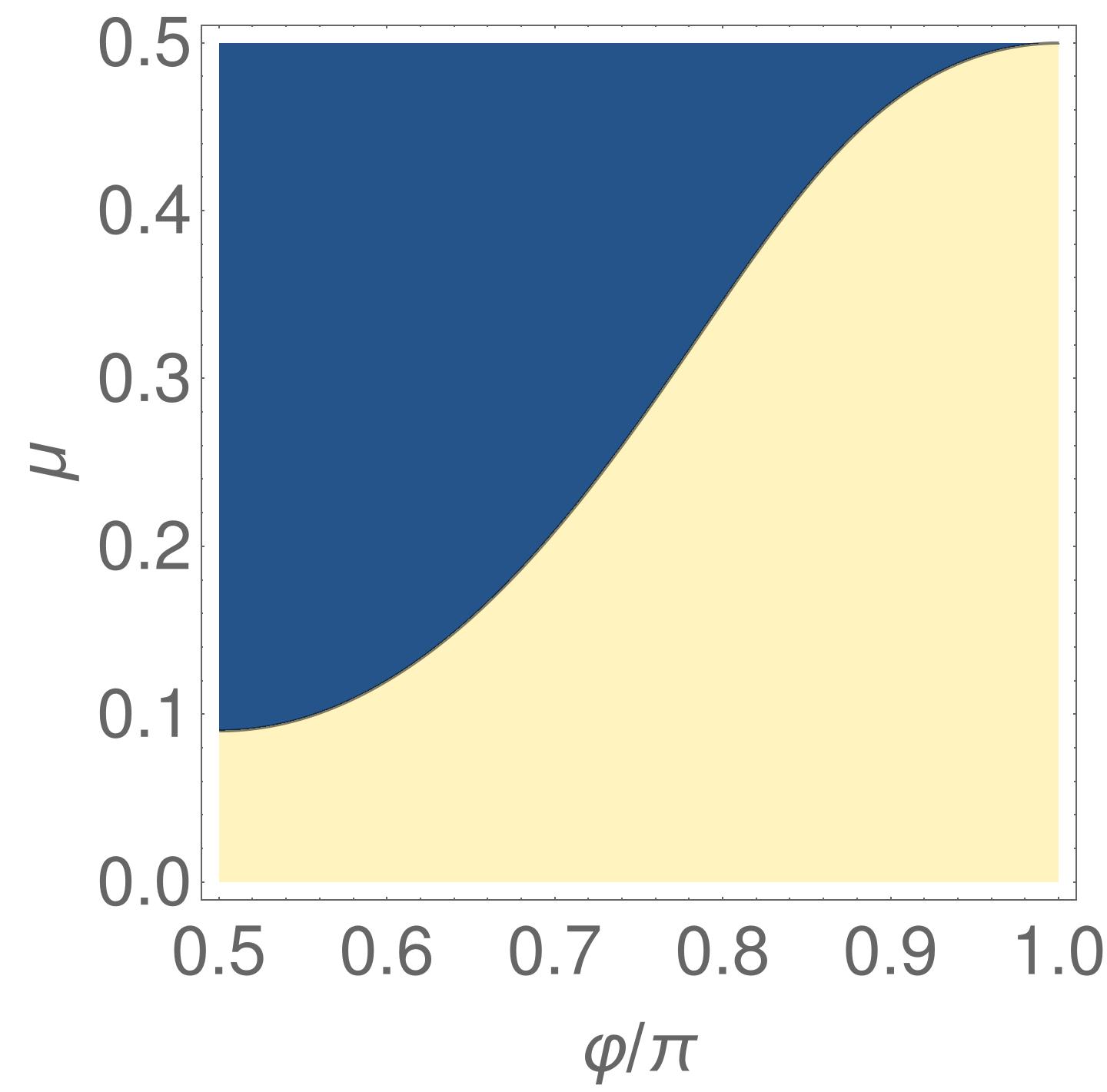}~\hfill
\includegraphics[width=0.33\textwidth]{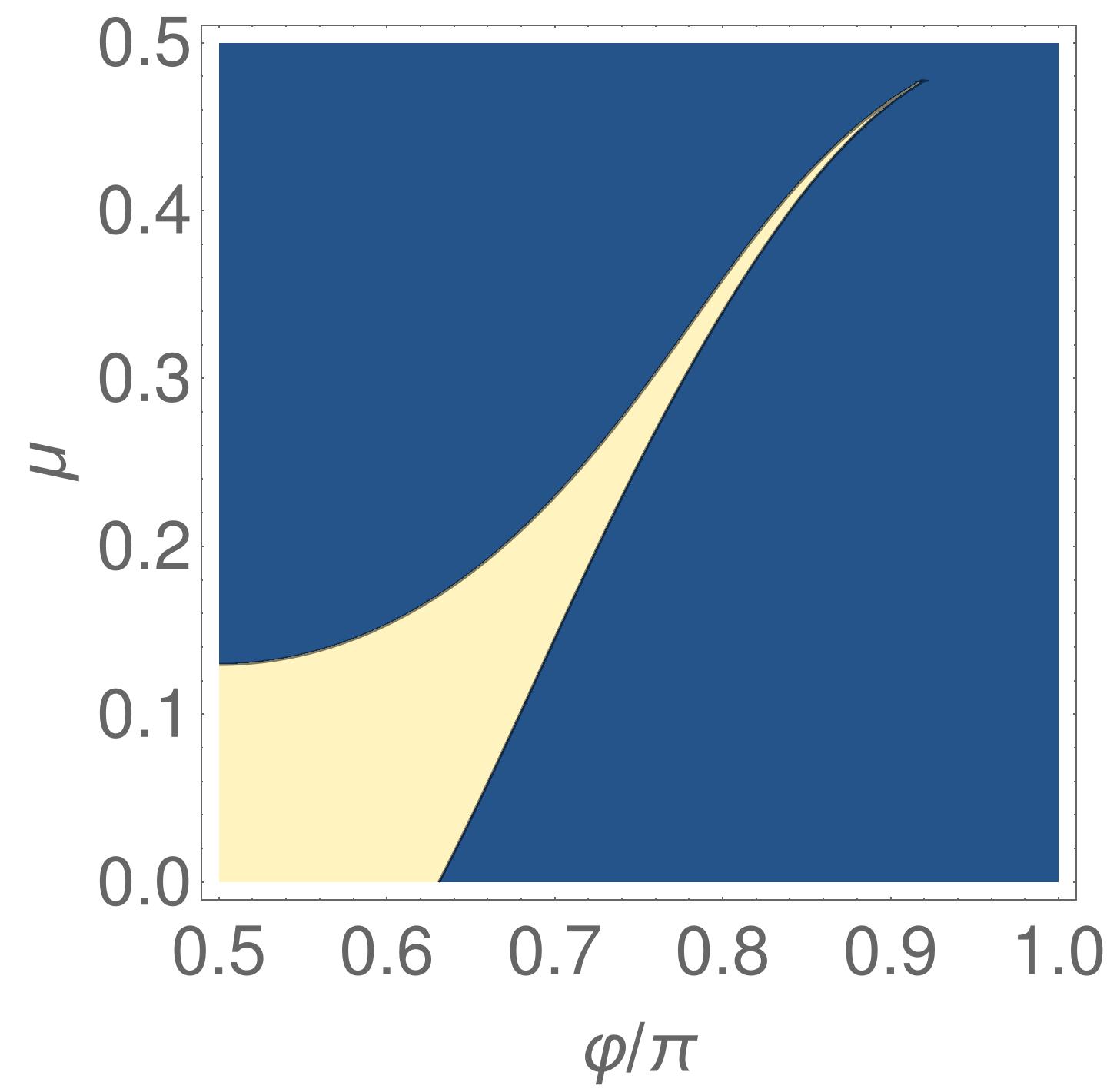}~
\caption{Hurwitz criterion for the asymmetric solution, Eqs.~(a)~\eqref{app:Hasym_2}, (b)~\eqref{app:Hasym_3} and (c)~\eqref{app:Hasym_4}. Yellow [blue] regions correspond to positive [negative] values. }\label{app:fig:Hcrit_asym}
\end{figure}

For the asymmetric solution, the calculation is more involved since we must use some trigonometric relations to replace all $\Theta$ dependencies.
First, from \eqref{app:14} we know
\begin{align}
&\tan(\Theta- \varphi) + \tan(\Theta + \varphi) = \frac{\tan \Theta}{\mu}~\label{app:15} \,, \\
&\tan^2(\Theta- \varphi) + \tan^2(\Theta + \varphi) = \frac{\tan^2 \Theta}{\mu^2} - 2 \tan(\Theta- \varphi) \tan(\Theta + \varphi)\,.\label{app:15_2}
\end{align}
Furthermore
\begin{align*}
\tan(\Theta- \varphi) + \tan(\Theta + \varphi) = \frac{2 \cos\Theta \sin \Theta}{\cos(\Theta + \varphi) \cos(\Theta - \varphi)}= \frac{2 \tan\Theta }{(1+ \tan^2 \Theta)\cos(\Theta + \varphi) \cos(\Theta - \varphi)}\,,
\end{align*}
and comparing with \eqref{app:15} it follows
\begin{equation}\label{eapp:16}
\cos(\Theta + \varphi) \cos(\Theta - \varphi) = \frac{2 \mu}{1+ \tan^2 \Theta}\,,
\end{equation}
which can then be used to obtain
\begin{align}\label{app:17}
\tan^2(\Theta- \varphi) + \tan^2(\Theta + \varphi) = \frac{\sin^2(2\Theta) +  \sin^2(2 \varphi)}{2 \cos^2(\Theta + \varphi) \cos^2(\Theta - \varphi)}= \frac{\left(\sin^2(2\Theta) +  \sin^2(2 \varphi)\right) \left(1+ \tan^2 \Theta \right)^2}{8 \mu^2}\,.
\end{align}
Now, inserting \eqref{app:17} into \eqref{app:15_2} we see that
\begin{equation}\label{app:18}
16 \mu^2 \tan(\Theta- \varphi) \tan(\Theta + \varphi) = 8 \tan^2 \Theta - \left(\sin^2(2\Theta) +  \sin^2(2 \varphi)\right) \left(1+ \tan^2 \Theta \right)^2\,.
\end{equation}
From \eqref{app:angle} it follows that
\begin{align*}
\sin^2(2\varphi)=&  4 \sin^2 \varphi \cos^2 \varphi  = 4 (1-2 \mu) \cos^2 \Theta \left(1 - (1-2 \mu) \cos^2 \Theta \right)\,, ~\\
\sin^2(2 \Theta)=& 4 \cos^2 \Theta \sin^2 \Theta =  4 \cos^2 \Theta (1 - \cos^2 \Theta)\,,
\end{align*}
which we insert into \eqref{app:18} and obtain
\begin{equation}\label{app:19}
\tan(\Theta- \varphi) \tan(\Theta + \varphi) = \frac{-1+2\mu + \tan^2 \Theta}{2 \mu}\,.
\end{equation}
Finally, using \eqref{app:15_2} we can replace any $\tan^2 \Theta$ via
\begin{equation}\label{app:20}
\tan^2 \Theta = \frac{1- \frac{\sin^2 \varphi}{1-2\mu}}{\frac{\sin^2 \varphi}{1-2\mu}} = \frac{1-2\mu}{\sin^2\varphi} -1\,.
\end{equation}
Next, using the replacements \eqref{app:15}, \eqref{app:15_2}, \eqref{app:19} and finally \eqref{app:20}, all coefficients of the Hurwitz criterion \eqref{app:criterion} are written as function depending solely on $\mu$ and $\varphi$:
\begin{align}
2 ( 1 - 2 \mu)>&0\,,~\label{app:Hasym_1} \\
4 ( 1 - 2 \mu) \mu \left(-1 + (1-2\mu)\csc^2 \varphi\right)>&0~\,, \label{app:Hasym_2} \\
\csc^2 \varphi\left(5 -28 \mu +40 \mu^2 - 16 \mu^3 - (-3 +2 \mu +4 \mu^2) \cos(2 \varphi) \right)>&0\,,~\label{app:Hasym_3} \\
-4 \mu\bigg( (4 \mu -1 )(1-5\mu + 8 \mu^2) + \left(3 -16 \mu +40 \mu^2 -16 \mu^3\right)\left(-1+(1-2\mu)\csc^2\varphi\right) \cr  + 3\left(4-7\mu + 4\mu^2\right)\left(-1+(1-2\mu)\csc^2 \varphi\right)^2\bigg)  > & 0\,.\label{app:Hasym_4}
\end{align}
Eq.~\eqref{app:Hasym_1} is always fulfilled for the considered $\mu$ values, Fig.~\ref{app:fig:Hcrit_asym} illustrates Eqs.~\eqref{app:Hasym_2}, \eqref{app:Hasym_3}, and \eqref{app:Hasym_4} in the panels~(a), (b), and (c), respectively. We see that Fig.~\ref{app:fig:Hcrit_asym}~(a) gives the pitchfork bifurcation to the symmetric stationary state and Fig.~\ref{app:fig:Hcrit_asym}~(b) does not apply taking  into account the existence interval $0\leq \mu \leq \frac12 \cos^2\varphi$. I.e.~the remaining Fig.~\ref{app:fig:Hcrit_asym}~(c) gives the lower stability border of the asymmetric solution. Rewriting the corresponding Eq.~\eqref{app:Hasym_4} we conclude that the asymmetric stationary state is stable if
\begin{align} 
&-( 4\mu-1)(1 - 5\mu + 8\mu^2) \sin^4 \varphi - 3(4 - 7\mu + 4\mu^2)( \cos^2 \varphi-2\mu)^2 \cr 
&-\Big(3 + 8\mu(\mu - 2)(1 - 2\mu)\Big)\left( \cos^2 \varphi-2\mu\right) \sin^2 \varphi > 0\,.
       \end{align} 
 The zero crossing of the left hand side indicates the locus of the secondary Hopf bifurcation  which renders the asymmetric state unstable.


%
\end{document}